\documentclass[letterpaper,twocolumn,11pt,unpublished]{quantumarticle}
\pdfoutput=1

\usepackage{amsfonts}
\usepackage{amssymb}
\usepackage{amsmath}
\usepackage{bbm}
\usepackage{graphicx}
\usepackage{verbatim}
\usepackage{hyperref}
\usepackage[english]{babel}
\usepackage[numbers]{natbib}
\usepackage[pdftex]{color}
\usepackage{dsfont}
\usepackage{wrapfig}
\usepackage[normalem]{ulem}
\usepackage{comment}
\usepackage[dvipsnames]{xcolor}

\usepackage{subfig}
\usepackage{tikz}
\usetikzlibrary{calc, shapes.geometric, arrows, decorations.markings}

\hypersetup{
    colorlinks=true,       % false: boxed links; true: colored links
    linkcolor=cyan,          % color of internal links
    citecolor=magenta,        % color of links to bibliography
    filecolor=magenta,      % color of file links
    urlcolor=cyan,           % color of external links
    runcolor=cyan
}

\graphicspath{{figs/}}

\newcommand{\ket}[1]{|#1\rangle}

\newcommand\rl[1]{\textcolor{purple}{#1}}

\begin{document}

\title{Towards solving the Fermi-Hubbard model via tailored quantum annealers}
%%%%%%% Alternatives? %%%%%%%%%%%%%%%%%%%%
% Simulating annealing dynamics of Fermi-Hubbard Models
% Proposal of quantum annealers tailored for simulating Fermi-Hubbard models: Numerical investigation and experimental perspectives 
% 

\author{Ryan Levy}
 \email{rlevy3@illinois.edu}
 \affiliation{Institute for Condensed Matter Theory and IQUIST and NCSA Center
for Artificial Intelligence Innovation and Department of Physics,
University of Illinois at Urbana-Champaign, IL 61801, USA}
 \affiliation{Quantum Artificial Intelligence Laboratory (QuAIL), NASA Ames Research Center, Moffett Field, CA, 94035, USA}
 \affiliation{USRA Research Institute for Advanced Computer Science (RIACS), Mountain View, CA, 94043, USA}
 
\author{Zoe Gonzalez Izquierdo}

\author{Zhihui Wang}

\author{Jeffrey Marshall}

\author{Joseph Barreto}
 \affiliation{Quantum Artificial Intelligence Laboratory (QuAIL), NASA Ames Research Center, Moffett Field, CA, 94035, USA}
 \affiliation{USRA Research Institute for Advanced Computer Science (RIACS), Mountain View, CA, 94043, USA}
 
\author{Louis Fry-Bouriaux}
 \affiliation{%
 London Centre for Nanotechnology, University College London, WC1H 0AH London, UK
}%

\author{Daniel T. O'Connor}
 \affiliation{%
 London Centre for Nanotechnology, University College London, WC1H 0AH London, UK
}%

\author{Paul A. Warburton}
 \affiliation{%
 London Centre for Nanotechnology, University College London, WC1H 0AH London, UK
}%
\affiliation{%
 Department of Electronic \& Electrical Engineering, University College London, WC1E 7JE London, UK
}%

% \author{Andrew J. Kerman}
% \affiliation{MIT Lincoln Laboratory, 244 Wood Street, Lexington, MA 02421}

 \author{Nathan Wiebe}
 \affiliation{ Department of Computer Science, University of Toronto, Toronto, ON M5S 2E4, Canada}
  \affiliation{ Pacific Northwest National Laboratory, Richland, WA 99352 , USA}

\author{Eleanor Rieffel}
 \affiliation{Quantum Artificial Intelligence Laboratory (QuAIL), NASA Ames Research Center, Moffett Field, CA, 94035, USA}

\author{Filip A. Wudarski}
 \email{filip.a.wudarski@nasa.gov}
 \affiliation{Quantum Artificial Intelligence Laboratory (QuAIL), NASA Ames Research Center, Moffett Field, CA, 94035, USA}
 \affiliation{USRA Research Institute for Advanced Computer Science (RIACS), Mountain View, CA, 94043, USA}

\begin{abstract}
    The Fermi-Hubbard model (FHM) on a two dimensional square lattice has long been an important testbed and target for simulating fermionic Hamiltonians on quantum hardware.  We present an alternative for quantum simulation of FHMs based on an adiabatic protocol that could be an attractive target for next generations of quantum annealers. Our results rely on a recently introduced low-weight encoding that allows the FHM to be expressed in terms of Pauli operators with locality of at most three.  We theoretically and numerically determine promising quantum annealing setups for both interacting 2D spinless and spinful systems, that enable to reach near the ground state solution with high fidelity for systems as big as $6\times 6$ (spinless) and $4\times 3$ (spinful). Moreover, we demonstrate the scaling properties of the minimal gap and analyze robustness of the protocol against control noise. Additionally, we identify and discuss basic experimental requirements to construct near term annealing hardware tailored to simulate these problems. 
    Finally, we perform a detailed resource estimation for the introduced adiabatic protocol, and discuss pros and cons of this approach relative to gate-based approaches for near-term platforms.
    %, we compare resource estimation for the introduced annealing protocol with gate based approaches for near term platforms.  %edit this

\end{abstract}

% \date{\today}

\maketitle

\section{Introduction}

Quantum computing is a novel theoretical computational paradigm that has entered the experimental phase with the first prototype devices being constructed \cite{Preskill2018,Johnson2011}. Through exploitation of quantum resources such as coherence and entanglement, quantum computation promises faster solution to certain problems \cite{Grover1996,Shor1997,Arute2019} than that offered by classical computers. One of the most prominent areas where scientists expect this computational advantage to happen is in modeling quantum chemical objects---atoms, molecules, and materials---for which superior complexity algorithms already exist \cite{Cao2019,Bauer2020}. However, these algorithms require sophisticated hardware implementations capable of running near fault-tolerant quantum computation with a prohibitive number of qubits (see for example \cite{kivlichan2020improved}). Therefore, one may expect to approach the material modeling problem from yet another perspective, namely to construct a single-purpose (application specific) device, that is tailored for computing a narrow class of problems. 
%Quantum annealers offered by the D-Wave Systems \cite{} belong to a single-purpose device class, where only problems expressible in the quadratic unconstrained binary optimization (QUBO) form can be executed on that hardware. 
Quantum annealers \cite{Finnila1994,Kadowaki1998,Fahri2001,Hauke2020} often belong to a single-purpose device class, where only problems expressible in the quadratic unconstrained binary optimization (QUBO) form can be encoded on the currently available hardware, as in those offered by D-Wave Systems \cite{dwave_website}.  
Mappings to QUBO of other Hamiltonians have been proposed--- such as for the electronic structure Hamiltonian \cite{xia2017electronic, streif2019solving}---however they require substantial overhead in the number of qubits (especially in the case of sparsely connected device architectures) and sophisticated data postprocessing, that allows the extraction of the relevant solution's signal.

In this paper we propose an annealing protocol that is tailored to arrive near-the-ground state (i.e. with expected energy lower than the first excited state) of the Fermi-Hubbard model (FHM), which is often used as a proxy playground for materials physics - in particular to study superconductive and insulating phases \cite{LeBlanc2015,Qin2021}.
% \jm{`near the gs'? do we mean gs and a few excited states? good to be more precise here, current statement unclear.}
Such a hypothetical single-purpose quantum machine could, in principle, be less prone to errors, since the controlling mechanism is restricted to a hardwired (hence more robust) set of instructions. 
Despite hardwiring, some tunability of the FHM parameters would still be required to fully access the ground state phase diagram and explore phase transitions, the ground state energy and potentially other observables (e.g. magnetization, spin density, superconductivity, etc.).

% \rl{more speculation} Additionally, it would be tunable to the degree that one can investigate a large arrays of FHM, thus to look carefully into phase transitions, system's energy and potentially other observables (e.g. magnetization, spin density, superconductivity, etc.). 

This contribution relies on a recently proposed compact fermion-to-qubit encoding \cite{Derby2021}, that provides means to map fermionic degrees of freedom (modes) into a qubit architecture. 
A fermion-to-qubit encoding is a crucial ingredient in most quantum algorithms that aim to solve fermionic problems. So far, a large number of methods exist \cite{Derby2021,Verstraete2005,Ball2005,jordanwigner1928,Bravyi2002,Setia2019,Steudtner2019,Jiang2019,Chen2022}, with arguably the most notable being the Jordan-Wigner transformation \cite{jordanwigner1928}. Each of the methods has its own pros and cons (see for example \cite{Clinton2020}); we also enumerate some differences in Table~\ref{tab:encoding} for comparison. Here we argue that the low weight (LW) encoding \cite{Derby2021} is well suited for annealing purposes, and we propose a protocol that is supported by numerical simulations and shows how to obtain a near-the-ground state outcome for varying system sizes. Additionally we analyze scaling properties in terms of minimal gap and robustness against control noise.  
%\jm{some difficulty with this also. you mean near the GS with polynomial scaling, or at a fixed size, or using exponential resources, or something else? and how near is near?} 
Since the LW encoding involves more physical qubits than the number of fermionic modes, it also introduces the notion of logical subspace (i.e. a subspace associated with solely fermionic modes), that is defined by the stabilizer formalism \cite{Gottesman1997}. 
Due to imposed symmetries, the proposed protocol operates in the logical space, provided that the evolution is coherent (not exposed to the effects of decoherence). 

The main motivation standing behind this protocol, is to provide a road map for the experimental realization of an annealing platform that can run FHMs. We deliberately restrict the analysis to the LW encoding and construct the entire evolution such that it requires couplers of order three or less. These couplers are currently under active investigation \cite{Leib2016,Chancellor2017,Melanson2019,Schondorf2019}, making this proposal experimentally feasible in the near future. Even though three-way couplers are not yet ubiquitous, we aim to theoretically convince the community that additional research endeavors towards development of these type of components would be beneficial in treating complex problems like FHMs, meaning it is an attractive alternative to other near term approaches, in particular variational algorithms run on gate based computers.

The remaining of the paper is organized as follows. In the next section we summarize the LW encoding, together with the associated stabilizer formalism. In Section~\ref{sec:protocol}, we present the annealing protocol with all its ingredients - drivers and schedule functions. The results are presented in Section~\ref{sec:results}, which include simulated ground state fidelity and observables. Additionally in Section~\ref{sec:experimental} we discuss other approaches to FHMs and compare them with the one we propose, as well as discussing its potential experimental implementations based on currently existing trends in superconducting qubits. Finally, we summarize our findings in Section~\ref{sec:conclusions} and provide an outlook on future directions, together with open problems that can bring us to constructing an annealing device capable of solving FHM.

\section{Low weight fermion-to-qubit mapping}\label{sec:LW}
% Description of DK encoding \cite{Derby2021} with the essentials that we need for this paper, i.e. square lattice case, stabilizers, starting with spinless and going into two layers for spinful.

% \fifi{Here I pasted what we've got from the other note}
% \zw{[[unify compact / LW encoding]]}

\begin{figure*}
    \centering
    \includegraphics[width = 0.95\textwidth]{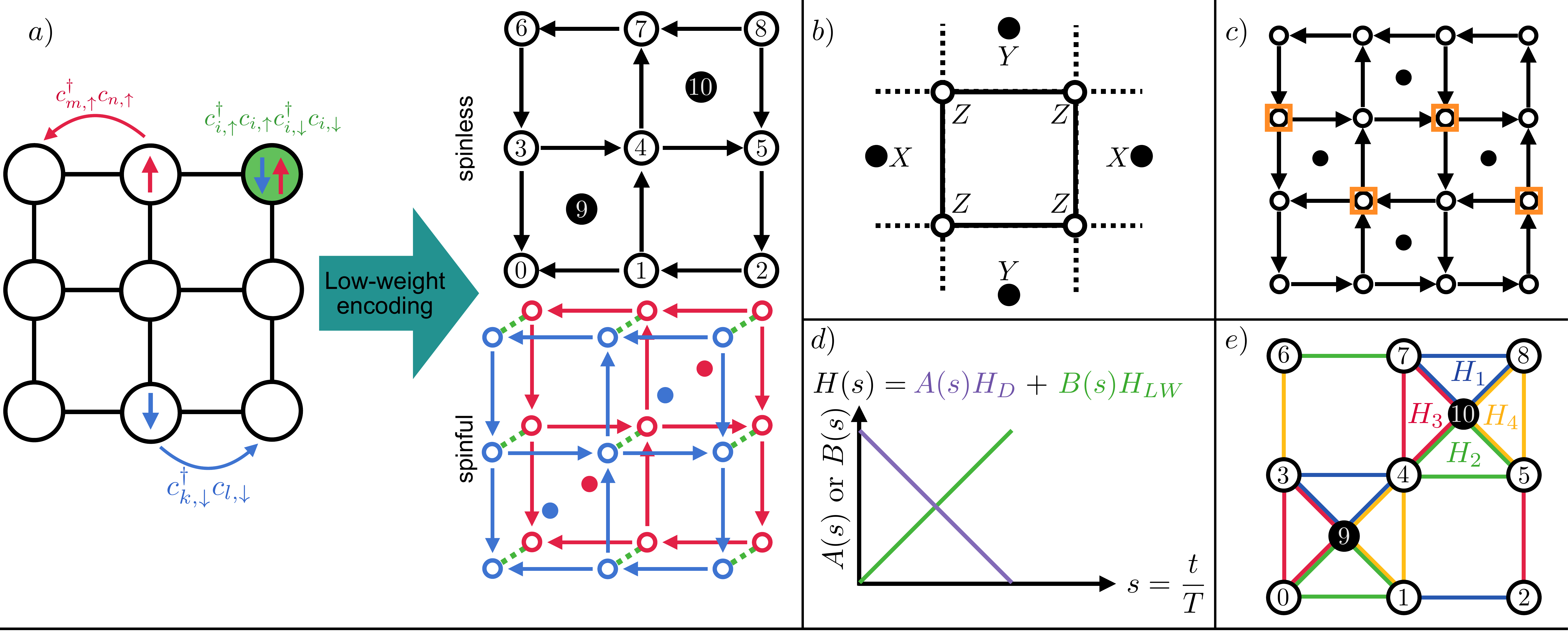}
    \caption{a) Left: Depiction of the Fermi-Hubbard model (FHM). Fermionic spins (red,blue arrows) can hop from site to site and experience an energy penalty $U$ (green) when located on the same site. Right: By encoding the FHM by the low-weight encoding we obtain two copies of a spinless lattice with auxiliary qubits. The $U$ interaction term involves interactions between sublattices, with green doted lines depicting the $U$ term between each site. 
    b) Example of a stabilizer for the low-weight encoding
    c) An example checkerboard pattern driver for $n_e=4$ fermions on a $4 \times 4$ square lattice with LW encoding encoding orientation denoted with arrows. The orange squares and white circles denote the $-Z_i$ and $+Z_i$ terms respectively to create the state $ c^\dagger_{10} c^\dagger_8 c^\dagger_7 c^\dagger_5\ket{0}^{\otimes N}$.
    d) Depiction of the linear annealing schedule, which smoothly interpolates between the initial driver Hamiltonian $H_D$ and the low-weight encoded problem Hamiltonian $H_{LW}$ over a time $T$. 
    e) The hopping terms of Eq.~(\ref{eq:hopping}) can be grouped into 4 non-commuting operators $H_1,\dots, H_4$, which have operators shown in different colors through the lattice. 
    }
    \label{fig:model_FH}
\end{figure*}

We briefly summarize the low weight encoding in Ref.~\cite{Derby2021} (see Fig.~\ref{fig:model_FH} for a schematic representation). Consider fermionic operators  $c_i^\dag, c_i$ that need to satisfy anti-commutation rules (here we drop the spin index)
\begin{eqnarray}
\{ c_i, c_j^\dag\} &=& \delta_{i,j},\\
\{c_i, c_j\} = \{ c_i^\dag,c_j^\dag\} &=& 0.
\end{eqnarray}
The LW encoding constructs operators involving (at most) 3-local Pauli terms to satisfy this fermionic algebra. 

Suppose that we have a $N_x \times N_y$ lattice with $N=N_xN_y$ sites that can be either unoccupied or occupied by a (spinless) fermion. We place a qubit at each vertex of the lattice along with $\leq 0.5N$ additional qubits (at the center of faces in a checkerboard pattern), denoted auxiliary qubits,  which aid in the LW encoding. 

With each vertex $V_j$ and edge $E_{i,j}$ we associate operators that are related to creation and annihilation operators and can be expressed in terms of Pauli operators
% \begin{equation}
% E_{i,j} = \left\{\begin{array}{ccl}X_i Y_j X_{f(i,j)} &  & \mathrm{if}\ (i,j)\ \textrm{is oriented downwards} \\-X_iY_jX_{f(i,j)} &   & \mathrm{if}\ (i,j)\ \textrm{is oriented upwards} \\X_iY_j Y_{f(i,j)} &   & \mathrm{if}\ (i,j)\ \textrm{is horizontal}\end{array}\right.,
% \end{equation}
{\small{
\begin{equation}
E_{i,j} = \left\{\begin{array}{cl}X_i Y_j X_{f(i,j)} &   \mathrm{if}\ (i,j)\ \textrm{is oriented downwards} \\-X_iY_jX_{f(i,j)} &    \mathrm{if}\ (i,j)\ \textrm{is oriented upwards} \\X_iY_j Y_{f(i,j)} &   \mathrm{if}\ (i,j)\ \textrm{is horizontal}\end{array}\right.
\end{equation}}}
where $f(i,j)$ is a function identifying the auxiliary qubit adjacent to the $(i,j)$ edge. When the $(i,j)$ edge has no nearby auxiliary qubit, the corresponding $X_{f(i,j)}$ or $Y_{f(i,j)}$ Pauli term is dropped. 
Vertex operators are $V_j=Z_j$. Creation and annihilation operators for FHM congregate into
\begin{eqnarray}
c_i^\dag c_j+ c_j^\dag c_i &=& -\frac{i}{2}\Big(E_{i,j}V_j+V_j E_{i,j}\Big), \label{eq:hopping}\\
c_{i,\uparrow}^\dag c_{i,\uparrow}c_{i,\downarrow}^\dag c_{i,\downarrow} &= &  \frac{1}{4}\big(I-V_{i,\uparrow}\big)\big(I-V_{i,\downarrow}\big),%\mathbbm{1}
\end{eqnarray}
which translates into Pauli operators as
\begin{eqnarray}
c_{i,\uparrow}^\dag c_{i,\uparrow}c_{i,\downarrow}^\dag c_{i,\downarrow} &\propto & \big(I-Z_{i,\uparrow}\big)\big(I-Z_{i,\downarrow}\big),\\ %\mathbbm{1}
% c_{i,\sigma}^\dag c_{j,\sigma}+ c_{j,\sigma}^\dag c_{i,\sigma} &\propto & X_{i,\sigma} X_{j,\sigma} Y_{f(i,j),\sigma}+ Y_{i,\sigma}Y_{j,\sigma}Y_{f(i,j),\sigma}\label{eq:horizontal}\\
% c_{i,\sigma}^\dag c_{j,\sigma}+ c_{j,\sigma}^\dag c_{i,\sigma} &\propto & X_{i,\sigma} X_{j,\sigma} X_{f(i,j),\sigma}+ Y_{i,\sigma}Y_{j,\sigma}X_{f(i,j),\sigma}\label{eq:vertical},
c_{i}^\dag c_{j}+ c_{j}^\dag c_{i} &\propto & X_{i} X_{j} Y_{f(i,j)}+ Y_{i}Y_{j}Y_{f(i,j)}\label{eq:horizontal}\\
c_{i}^\dag c_{j}+ c_{j}^\dag c_{i} &\propto & X_{i} X_{j} X_{f(i,j)}+ Y_{i}Y_{j}X_{f(i,j)}\label{eq:vertical},
\end{eqnarray}
where Eq.~(\ref{eq:horizontal}) describes edges aligned horizontally, while Eq.~(\ref{eq:vertical}) corresponds to vertical edges. Thus an interacting FHM Hamiltonian will have terms at most 3-local of the form $XXX, XXY, YYY$, $YYX$, and connectivity of at most $6+1$ (for square lattices), with the addition of the $ZZ$ interaction term. We provide an explicit enumeration of the encoded Hamiltonian in Appendix~\ref{sec:app_example}).

\subsection{Stabilizers}

With the introduction of auxiliary qubits, the Hilbert space of the encoded Hamiltonian is now larger than that of the original problem. To fix this, and ensure the encoding admits a proper fermionic algebra, a set of stabilizers is required \cite{Derby2021}. 
We differentiate between \textit{logical} states, which refers to states which have $+1$ values for all possible stabilizers measurements, and \textit{non-logical} states, all other non-physical states. 
%%\footnote{Or all loops of edge operators must be in the $+1$ eigenspace}

We can iterate all stabilizers on a given graph. For a sequence $p$ of $k=|p|-1$ sites around a face $p=\{p_1,\dots,p_{k+1}=p_1\}$, the stabilizer is given by  
\begin{equation}
    i^{k} \prod_{i=1}^k E_{p_i} E_{p_{i+1}} = \mathbb{I}.
\end{equation}
Each odd face (those with auxiliary qubits) has a trivial stabilizer of the identity, while each even face (those without) have a stabilizer of at most Pauli weight 8 on a square lattice: $Z$ terms around the face, $X$ terms on the left/right auxiliary qubits and $Y$ terms on the up/down auxiliary qubits (see Fig.~\ref{fig:model_FH}). %, shown in Fig.~\ref{fig:Stabilizer}. 

Note that there are two special cases of the LW encoding where after the fermionic Hamiltonian is projected into the positive stabilizer subspace, there will be a different number of states than that of the expected FHM Hamiltonian. If there is one more odd face than even (more faces with an auxiliary qubit), then the only states represented will be those with even numbers of fermions. If there are more even faces than odd (more faces without an auxiliary qubit) then there will be a free auxiliary degree of freedom, leading to a degeneracy for every state. For lattices that are even in both $N_x$ and $N_y$, we always adopt the former choice such that there is a minimal number of overall qubits to represent the lattice. 

Finally we remark on the spectrum of the LW encoded Hamiltonian. 
%Empirically we find that while there are equal numbers of logical and non-logical states, non-logical states are always lower in energy to their logical counterparts. 
There are generally $2^{N/2}$ more non-logical states than logical, and as the non-logical states do not properly encode the fermionic signs, they occur lower than their logical counterparts \cite{Spencer2012}. 
Frequently, and more commonly as the system grows larger, there are many of these non-logical states between the non-logical ``ground state'' of the encoded Hamiltonian and the logical ground state (an excited state in the encoded spectrum).

\section{Annealing protocol}\label{sec:protocol}
Now we plan to utilize the encoded fermionic Hamiltonians and find their (logical) ground states via an annealing protocol. We set the annealing protocol problem Hamiltonian to be a fermionic Hamiltonian of LW encoding $H_{LW}$ (either spinless or FHM), alongside a driver Hamiltonian $H_D$ and time schedule. These components come together in the annealing of a time dependent Hamiltonian 
% We assume the problem Hamiltonian to be of a fermionic Hamiltonian (either spinless or FHM), with an additional need of: i) a driver Hamiltonian, and ii) a time schedule

\begin{equation}\label{eq:annealing}
    H(s) = A(s) H_D + B(s) H_{LW},
\end{equation}
for $s= \frac{t}{T}$, with total annealing time $T$. Above we denote $H_D$ the driver, and $A(s), B(s)$ are schedule functions which control the interplay between the two Hamiltonians. Conventionally, one tries to achieve close to adiabatic evolution, by starting in the ground state of $H_D$ and with $A(0)\gg B(0) \approx 0$ and then slowly increasing $B(s)$ while simultaneously decreasing $A(s)$, such that at $s=1$ the driver is turned off - $B(1)\gg A(1) \approx 0$. The profile of the time-dependent functions $A(s), B(s)$ and the rate of switching them on and off, will determine if the evolution satisfies the adiabatic condition. The adiabatic theorem guarantees ending in the ground state of the problem Hamiltonian $H_{LW}$, which is equivalent of finding the solution to the considered problem, when the evolution is performed slowly enough~\cite{Albash2018}. 
% Note, that at this level we allow schedule functions to take arbitrary shape (as long as the time evolution in Eq.~\eqref{eq:annealing} is close to adiabatic). 
Note, so far we have not made any assumptions on the shape of the annealing schedule functions (as long as the time evolution in Eq.~\eqref{eq:annealing} is close to adiabatic).
However, in real experiments there are additional constraints that arise due to hardware limitations which can exclude certain type of switching functions. Therefore, for theoretical ease we will limit ourselves to the linear schedule functions, that is $A(s) = 1-s$ and $B(s) = s$, with $s=t/T$ directly controllable by the total annealing time $T$. 
We explore the robustness of our approach to alternative schedules in Appendix~\ref{sec:app_schedules}. 
% \rl{Maybe can refer to discussion on experimental perspectives also if it has the info.}

% \fifix{Here we need to point out that the easy to prepare part is only on the main grid points, while the auxiliary qubits need also to live in the logical subspace, and this part might be non-trivial, which we don't fully explore in this paper and assume access to it. We may consider a special appendix on this problem to properly define the problem} 

When choosing a driver Hamiltonian, its ground state, i.e. the initial state, must be prepared directly by near-term hardware, which suggests that the driver should be of low local form, otherwise the control is less reliable. However, the low-weight encoding prevents us from easily using a standard transverse-field driver. 
This is because the stabilizer formalism creates non-logical states with energy below the target ground state in the problem Hamiltonian. Thus the ground state of the full problem Hamiltonian may have lower energy than the logical fermionic ground state of the equivalent Jordan-Wigner encoded model. 

In order to address this problem, we impose extra conditions on the driver Hamiltonian, such that its symmetries guarantee that we end up in the logical subspace (assuming coherent evolution, and no symmetry breaking noise). This is achieved when $H_D$, and hence also $H(s)$ commute with all stabilizers of the encoding, i.e. $[H_D, S_k]= [H(s), S_k] = 0$ $\forall k$ denoting different stabilizers of the encoding. 

The two other alternatives beyond this approach are: i) add a penalty term to the Hamiltonian, composed of all the stabilizers, to shift the logical states energetically lower than the non-logical states, ii)  project the LW Hamiltonian to the logical subspace and anneal it. Neither of these options is viable for near term implementation, since they require access to more sophisticated hardware components than Pauli weight three - in case of i) one would need to employ Pauli weight 8 terms, while for ii) the non-locality is increased due to projection method, which also destroys compact form of the original couplers. Hence we will restrict to the vanilla LW Hamiltonian and impose symmetries by the driver's form. 

% \jm{my first thought at addressing the issue was to add a penalization term to the states outside the stab +1 eigenspace (so the logical GS is the true GS). i think the argument against this is that it would be non-local. could be worth mentioning.}

% \rl{Ensuring this condition, in our case, means that an initial state will generally have a non-trivial entangled state related to a (planar) toric code living on the auxiliary qubits. The preparation of this initial entangled state is assumed to be given (e.g. through an oracular access). More detailed analysis is left for future research direction, here we only restrict the discussion to potential ways of creating such an oracle (see Appendix~\ref{sec:app_stateprep}).}

This extra condition combined with the need of a low local form, in our case, means that an initial state should be a product state in the bulk, and non-trivial entangled state related to a (planar) toric code prepared on the auxiliary qubits. We mainly focus on the former part, while the latter is assumed to be given (e.g. through an oracular access). We discuss potential ways of creating such an oracle in Appendix~\ref{sec:app_stateprep}, while a full analysis of the problem is left for future work.

%More detailed analysis of the latter is left for future work, here we only restrict the discussion to potential ways of creating such an oracle (see Appendix~\ref{sec:app_stateprep}).

% Requiring this condition on the driver does not ensure an initial state preparation protocol that may be easy to implement, however. For drivers that are local in the bulk, the auxiliary qubits must be prepared into a non-trivial state. This is not fully explored in this work, but we elaborate on potential solutions in Appendix~\ref{sec:app_stateprep} and assume access to a logical initial state. \rl{Maybe could elaborate potential solutions in the experimental perspectives section (subsections?) rather than appendix? It does depend on what emphasis we want to put on experimental feasibility.}

Additionally, we constrain our driver to commute with the particle number operator $N_e = \sum_i N_i$, i.e. $[H(s), N_e]=0$, such that the  evolution preserves the total number of particles in the system. At first glance, this extra symmetry may be perceived as a drawback, but in fact it allows us to explore spectral properties of the encoded FHM in more detail. In particular, one can evolve to the lowest energetic state with a fixed particle number, and based on that parameter adjust the difficulty level of the computational problem. Furthermore, the particle number operator acts solely on the bulk, while leaving auxiliary qubits untouched, making the entire construction more intuitive and straightforward.

Finally, we comment on a measurement protocol after the annealing is completed. At minimum, one needs 4+1 (4 hopping, 1 interacting) independent Pauli string measurements to obtain the energy \cite{Clinton2020}, corresponding to 4 measurements of all Pauli X or Y with different combinations of X and Y on the auxiliary qubits, plus an additional measurement of every qubit in the Pauli-Z basis to obtain density and spin information. In this work we focus exclusively on the noiseless and coherent evolution. However, in the case of a noisy environment causing leakage to the non-logical subspace, one may still retrieve logical information through a set of suitably chosen measurements. By measuring additional observables, either combinations of stabilizers and energy terms using a subspace expansion \cite{McClean2020} or random measurements utilizing a logical classical shadow routine \cite{HongYe2022}, energy measurements can be projected back into a logical subspace. 

% One potential method of achieving this is described in  \cite{McClean2020} where in addition to the standard Pauli string measurements, stabilizers are measured simultaneously. 

\subsection{Drivers}

% \begin{figure}
%     \centering
%     \includegraphics[width=0.5\columnwidth]{4x4_checkerboard.pdf}
%     \caption{An example checkerboard pattern driver for $n_e=4$ fermions on a $4 \times 4$ square lattice with LW encoding encoding orientation denoted with arrows. The red squares and blue circles denote the $-Z_i$ and $+Z_i$ terms respectively to create the state $ c^\dagger_{10} c^\dagger_8 c^\dagger_7 c^\dagger_5\ket{0}^{\otimes N}$. We order the lattice row by row, left to right, starting from the bottom left corner.   }
%     \label{fig:checkerboard_driver}
% \end{figure}

% \fifi{Not sure if that should be a separate subsection, but if we keep it as it is, then here we would describe properties and scaling of drivers, starting with generic ones (satisfying commutation relationships), then narrowing down to checkerboard patter. The generic mathematical description should be augmented by some illustrative graphic examples (for ease of reading). }
Drivers $H_D$ that are of low local form, and satisfy commutation relationships $[H(s), S_k] = [H(s), N_e] = 0$ and $[H(s),H_D]\not=0$ can be constructed by placing $\pm Z_i$ at the $i$-th vertex such that a product state is formed in the Pauli-$Z$ basis, or
\begin{align}
    H_D = -\sum \pm Z_i,
    \label{eq:driver}
\end{align} where the number of the $-$ signs provides a method to target particle number. 
To ensure this state is logical, the auxiliary qubits must be prepared in the proper state, analogous to forming a toric-code ground state via either an efficient state preparation scheme \cite{Higgott2021} or the (up to) Pauli weight 4 terms of the stabilizer with the appropriate sign to ensure a +1 measurement. The latter method we call having ``all'' driver auxiliary terms included. These high weight terms in fact commute with a large number of Hamiltonian terms, and so we find frequently that including them in the driver does not necessarily yield better performance (see next section). We will assume generally, that the ground state to Eq.~\eqref{eq:driver} has been prepared in a logical state (see Appendix~\ref{sec:app_stateprep}).  
% \fifix{if we write appendix of state prep, we may refer here to it for more details} 

For each particle number sector of $n_e$ fermions, we can distinguish an exponentially large number of drivers of this form. We identify a polynomial subset in which we observe good performance,  we call them the \textit{checkerboard pattern drivers}. These drivers are formed by placing alternative $+$ and $-$ signs such that the ground state forms a checkerboard pattern, one example shown in Fig.~\ref{fig:model_FH}c. In the special case of $n_e=2$ fermions, we place 1 fermion per row. 
% according to the following prescription: i) on the $i$-th vertex qubit place $\pm Z_i$, and ii) on $j$-th auxiliary qubits place $X_j$ or $Y_j$ (depending on the lattice structure).  

\begin{table}[ht]
    \centering
    \begin{tabular}{lccc}
     &   JW & VC &   LW\tabularnewline
      
    \hline 
    \hline 
    Reference &   \cite{jordanwigner1928} &  \cite{Verstraete2005,Ball2005} &  \cite{Derby2021}\tabularnewline
    Total Qubits &   $L^{2}$ & $2L^{2}$ &  $1.5L^{2}-L\pm1$\tabularnewline
% \multicolumn{1}{l}{$1.5L^{2}-L\pm1$}\tabularnewline
% & & & $-L\pm1$ \tabularnewline
    % Max PW &   $O(L)$ & $4$ &   $3$\tabularnewline
        % Max& & & \tabularnewline
        Pauli Weight &   $O(L)$ & $4$ &   $3$\tabularnewline
    \end{tabular}
    \caption{Comparison of total number of qubits and maximal Pauli weight for Jordan-Wigner (JW), Verstraete-Cirac-Ball (VC), and Low Weight (LW) encodings of a FHM on a $L\times L$ square lattice.  
    %We consider analogous drivers of the form eq.~\ref{eq:driver}. 
    }
    \label{tab:encoding}
\end{table}

\begin{figure}
    \centering
    \includegraphics[width=\columnwidth]{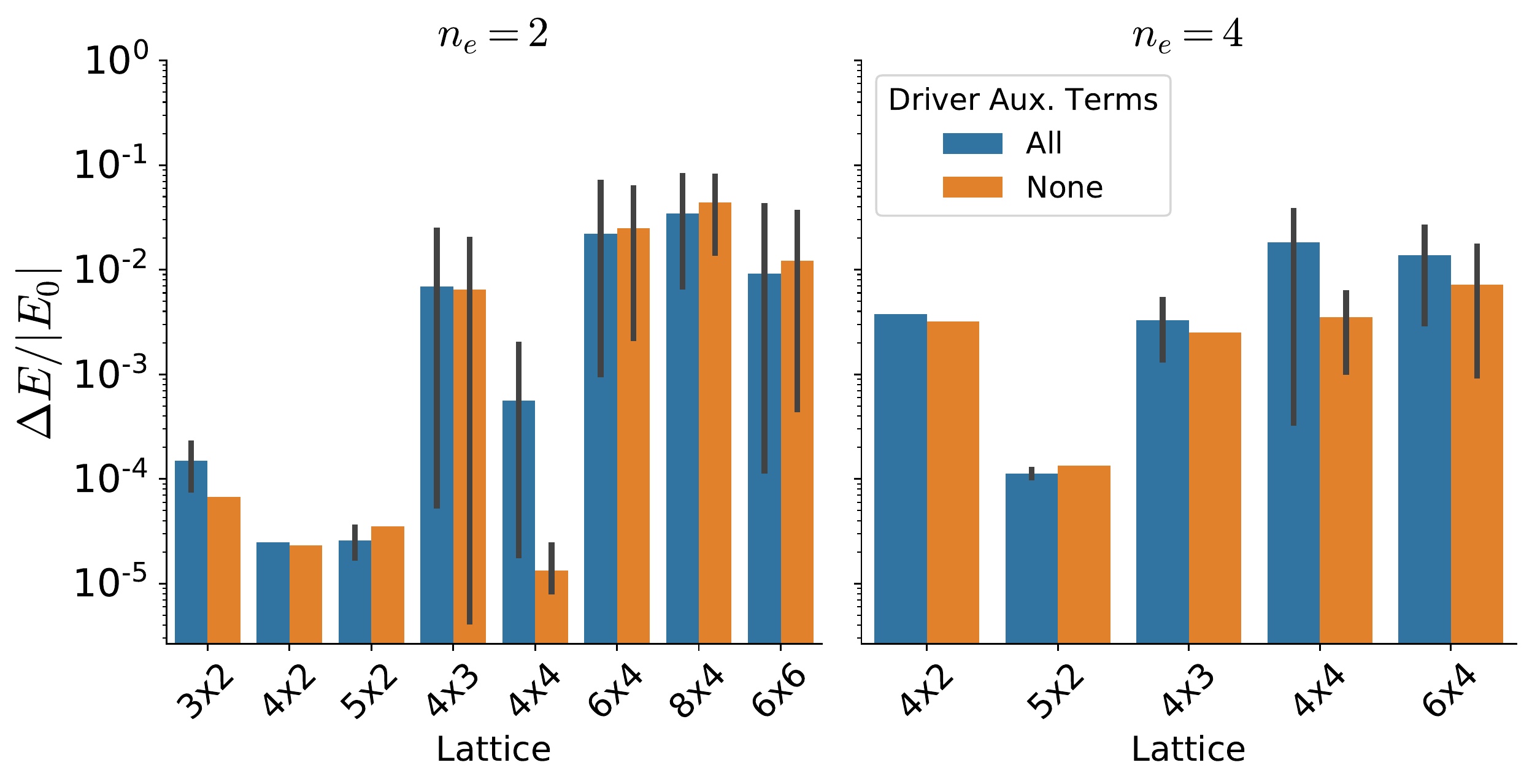}
    \caption{Relative ground state energy error $\Delta E/|E_0|$ with interaction strength $V/t=1$ for a given lattice shape for $n_e=2$ and $n_e=4$ spinless fermions using checkerboard drivers. Annealing simulation was performed with a linear schedule and total time $T=100$ in arbitrary units.
    The bars represent the mean $\Delta E/|E_0|$ error and error bars represent the maximum range of errors for all given checkerboard drivers, and we differentiate between an entirely local (Pauli weight 1) driver with no terms on the auxilary qubits or a driver with all terms on the auxilary qubits (up to Pauli weight 4). Drivers for $n_e=2$ additionally have no terms with an initial state containing a fermion in a corner of the lattice.   }
    \label{fig:spinless_deltaE}
\end{figure}

\section{Results}\label{sec:results}
All annealing simulations are noise-free, performed using the QuTiP library \cite{qutip1,qutip2} using the unitary Schr{\"o}dinger equation solver. The FHM is encoded on a square or rectangular lattice with open boundary conditions. We use a total annealing time of $T=100$ (in arbitrary time units with $\hbar=1$) and work in the subspace of a given particle number and logical states only. 
% \jm{by arbitrary, you mean units of the energy scale with $\hbar=1$? you want to make sure someone reading this could in principle reproduce the results, so good to be clear what is meant here...}

First we will benchmark our approach using spinless models where we can access large system sizes, looking at even fermion particle number. Then we will turn to a spinful Fermi-Hubbard model which due to computational complexity we can only access up to a $4\times 3$ lattice. 
Simulations are done with even numbers of fermions at or below half-filling. 

\subsection{Spinless models}
% \jm{minor, but i notice $t$ is used above e.g fig 1 and text to represent time ($s=t/T$). may consider changing notation.}
We first benchmark our approach using a spinless fermion system, a $t$-$V$ model
\begin{align}
    H_{LW} = -t\sum_{\langle ij\rangle} c^\dagger_ic_j + h.c. + V\sum_{\langle ij\rangle} n_i n_j,
\end{align}
where $\langle ij\rangle$ denote nearest neighbors with open boundary conditions. We fix $V/t=1$. Note that the hopping $t$ (not to be confused with $t$ used for time in the previous section) term will follow eqs.~\eqref{eq:horizontal},\eqref{eq:vertical} and the $V$ term will be $n_i n_j \sim (I-Z)_i (I-Z)_j$ (we explicitly construct an example of the encoding and driver in Appendix~\ref{sec:app_example}). The $V$ interaction term was chosen to mimic the $U$ interaction term of the FHM.

\begin{figure}
    \centering
    \includegraphics[width=\columnwidth]{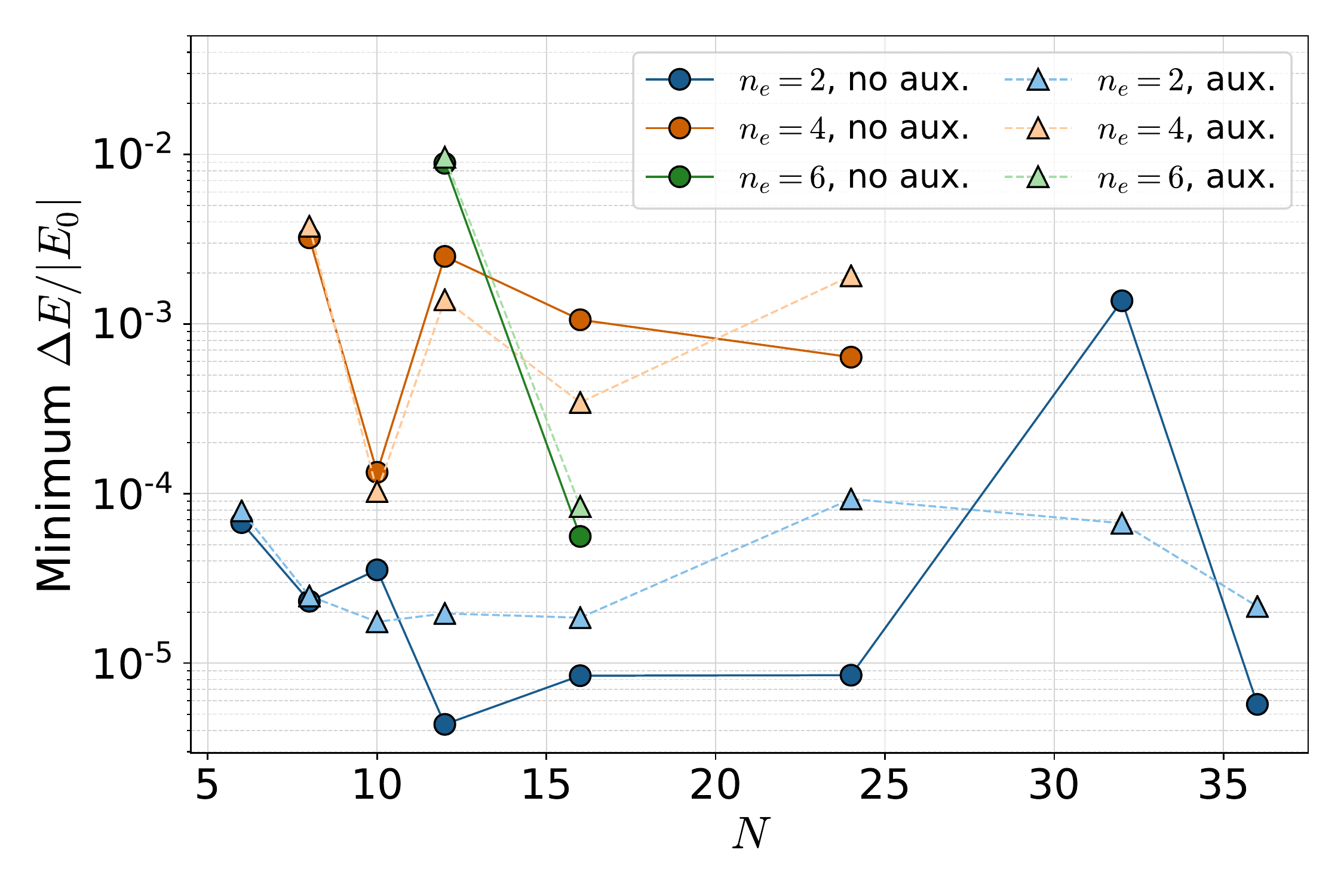}
    \caption{
    %Maximum fidelity out of all checkerboard pattern drivers, between the annealed state and the true $H_{LW}$ ground state at $V/t=1$. 
    Minimum $\Delta E/|E_0|$ from Fig.~\ref{fig:spinless_deltaE} for various spinless fermion lattices of $N$ points (excluding auxiliary qubits) at $V/t=1$.
    Annealing simulation was performed with a linear schedule and total time $T=100$ in arbitrary units. Drivers for $n_e=2$ additionally have no terms with an initial state containing a fermion in a corner of the lattice.}
    \label{fig:spinless_max_fid}
\end{figure}

We evolve the Hamiltonian $H(s)$ in Eq.~\eqref{eq:annealing} using the $t - V$ model as the problem Hamiltonian. Simulating the annealing, we obtain a final annealed state and compare its properties to the true ground state of the model (which can be obtained via exact diagonalization techniques). Specifically we observe the difference in energy, $\Delta E = \langle H_{LW} \rangle - E_0$, where $E_0$ is the logical ground state energy, and $\langle H_{LW} \rangle$ is the expectation value of the LW Hamiltonian at the end of annealing. 

By iterating over all checkerboard pattern drivers, in Fig.~\ref{fig:spinless_deltaE} we find fairly low errors in energy in both small and large lattice sizes. When considering to use a local driver (``None") over a driver which has terms on the auxilary qubits (``All"), there seems to be in general either a small increase to the energy or in a few cases an improvement to a lower average energy. We believe this is partly due to the  commutation of these auxiliary driver terms with many terms in the Hamiltonian, especially those that are in the same location.  
% \jm{here or in fig define $\Delta E$? I assume $\langle H_{LW}\rangle - E_0$?}

Note that some lattice sizes, such as $4\times 3$, have significant error bar outliers partially due to the lack of `bulk' to place the initial fermions away from the edges. In that specific lattice, of the 5 unique local checkerboard drivers, 4/5 have $\Delta E/|E_0| < 10^{-2}$ while only one has $\Delta E/|E_0| \approx  0.02$.
As system size increases the ability to place these initial fermions in the bulk of the system, helps make the annealing energy error more consistent. Using two particles, we explore the correlations on how to construct the $n_e=2$ driver in Appendix~\ref{sec:app_tw_corr}. In addition, for larger number of particles, there are relatively few (if not only one) checkerboard patterns and the error range is relatively consistent across system sizes. 

One concern, however, would be as the lattice size increase both the number of stabilizers and thus Pauli weight 4 auxiliary terms in the driver increase. Counting stabilizers for a $L\times L$ lattice there will be $(L-1)^2/2$ stabilizers, where all non-edge stabilizers are Pauli weight 4. For example, a $4\times 3$ lattice has 2 stabilizers of which are at most Pauli weight 3, while a  $6\times 6$ lattice will have 13 stabilizers, requiring 8 Pauli weight 4 terms. However, when removing these terms at fairly large lattice sizes for $n_e=2$, we see little to no change in performance. 
% \fifi{This we associated with the fact, that the auxiliary qubits are mainly responsible for encoding fermionic commutation rule, while the bulk qubits determine energetic properties of the system.} \fifix{Can we say that?}

Finally we compute the relative energy error for the best checkerboard driver (i.e. minimal $\Delta E/|E_0|$) for each lattice as a function of total lattice points $N$ in Fig.~\ref{fig:spinless_max_fid}. For dilute doping $n_e=2$, we find very low $\Delta E/|E_0|$  even out to large system sizes. When comparing rectangular to square lattices, the ladder like lattices have larger energy variations than the square lattices. This appears to arise from the rectangular lattices having a stronger initial location dependence than the square lattices (see Appendix~\ref{sec:app_tw_corr}). Considering a larger number of electrons we see similar low energy results, for intermediate system sizes not exceeding $\Delta E/|E_0|=10^{-2}$.   
%Larger number of electrons we see similar results, with the lowest fidelity for $n_e=4,6$ being $\approx 0.81$.  

% \jm{would a plot here or below of e.g $\Delta E$ vs annealing time $T$ be useful? how does the minimum gap scale with system size, empirically, for various drivers? (didnt read the rest yet, so maybe it's already there). I assume the ones that do well have the larger gap?}

\begin{figure}[t]
    \centering
        \subfloat[]{
      \includegraphics[width=\columnwidth]{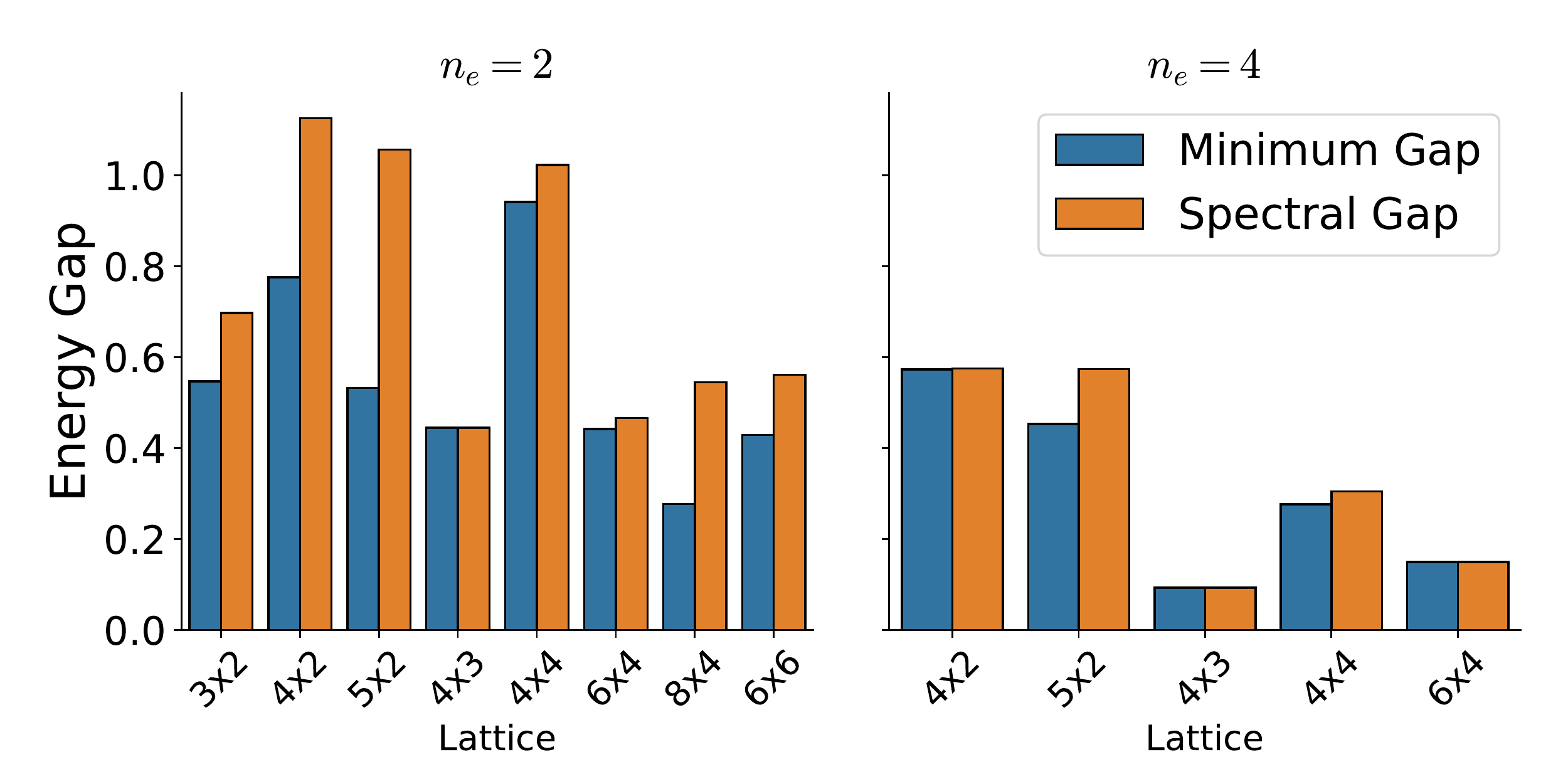}
    \label{fig:gap_abs_spinless}
    }\\
    \subfloat[]{
      \includegraphics[width=\columnwidth]{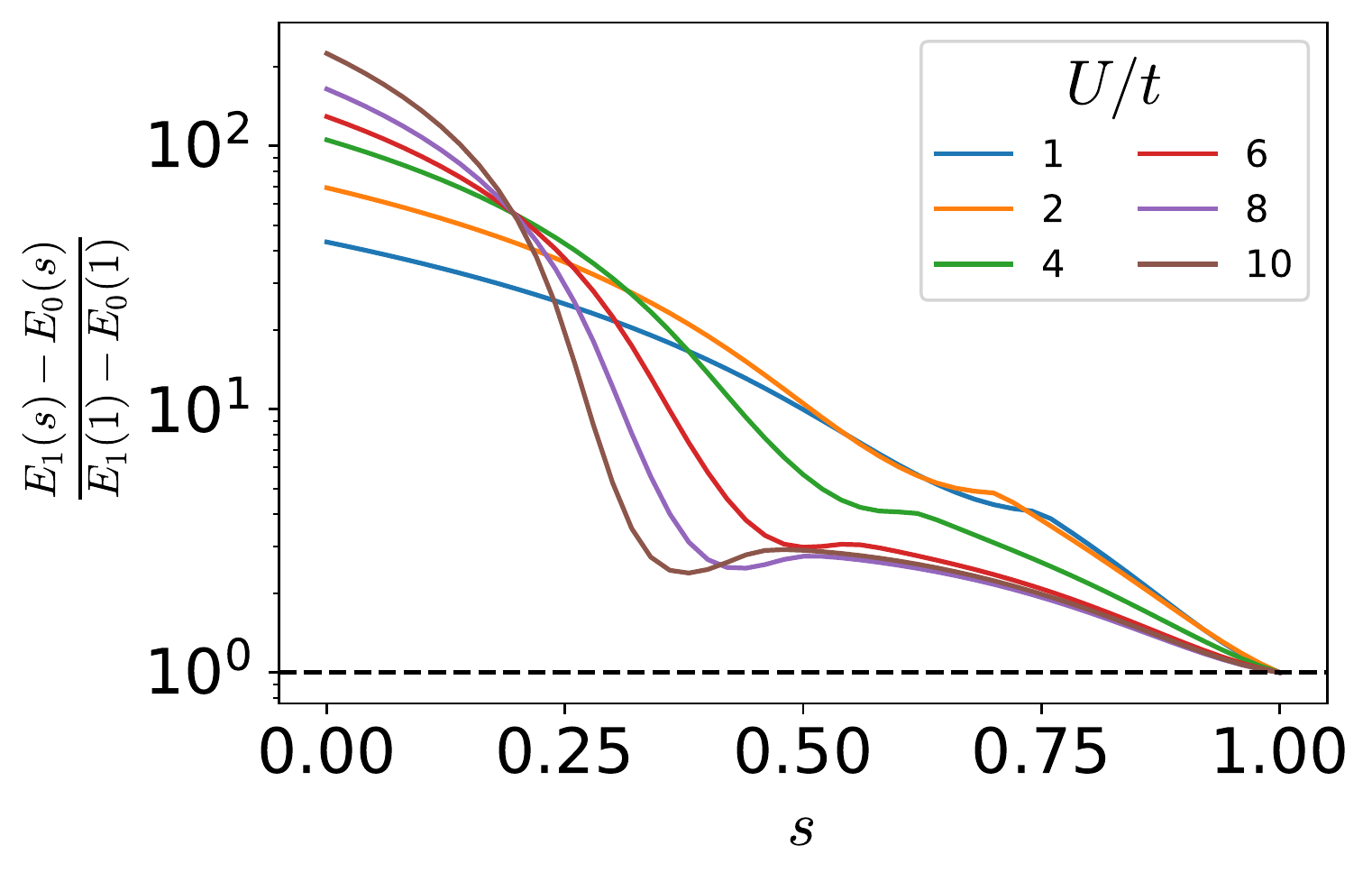}
    \label{fig:gap_hubbard}
    }
    \caption{
    a) absolute minimum gap $\min_s [E_1(s)-E_0(s)]$ for various lattice sizes and number of electrons alongside the spectral gap from (a). The drivers correspond to those in Fig.~\ref{fig:spinless_deltaE}. 
     b) relative spectral gap as a function of total anneal $s = \frac{t}{T}$ for the $4\times 3$ Hubbard model at $n=2/3$ doping. The drivers correspond to those in Fig.~\ref{fig:hubbard_max_fid_same} i.e. matching spin drivers. The minimum gap occurs at the end of the anneal, corresponding to the spectral gap.
     }
    \label{fig:gap_all}
\end{figure}

We can additionally analyze the time dependent energy levels $E_i(s)$ of the annealing problems performed in Fig.~\ref{fig:spinless_deltaE}. To find these levels, we perform exact diagonalization on $H(s)$ for 1000 equally distributed points $s\in [0,1]$. We can compare the spectral gap for these systems, $E_1 - E_0$, to the minimum gap in Fig.~\ref{fig:gap_abs_spinless}. Although there is some decay as system sizes increase, the gap appears to track the spectral gap with at least 50\% of the spectral gap value.  Note that there is no efficient generic algorithm to determine if there is a spectral gap for a model in general \cite{Cubitt2015}, and thus annealing can be used as a potential probe toward understanding the gap.

\subsection{Hubbard (Spinful) model}

\begin{figure}
    \centering
    \subfloat[Matching spin drivers]{
       \includegraphics[width=\columnwidth]{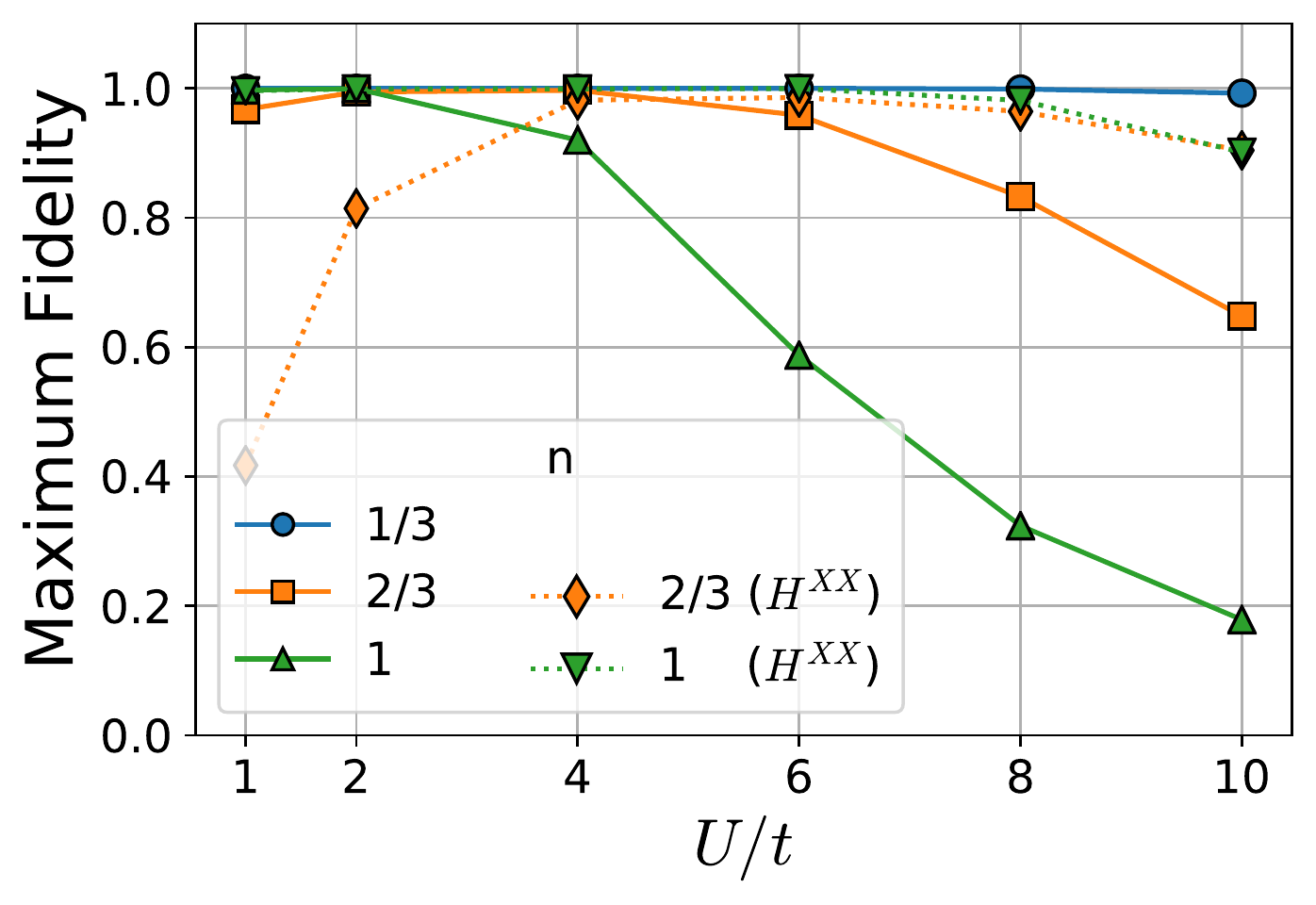}
    \label{fig:hubbard_max_fid_same}
    }\\
        \subfloat[Different spin drivers]{
       \includegraphics[width=\columnwidth]{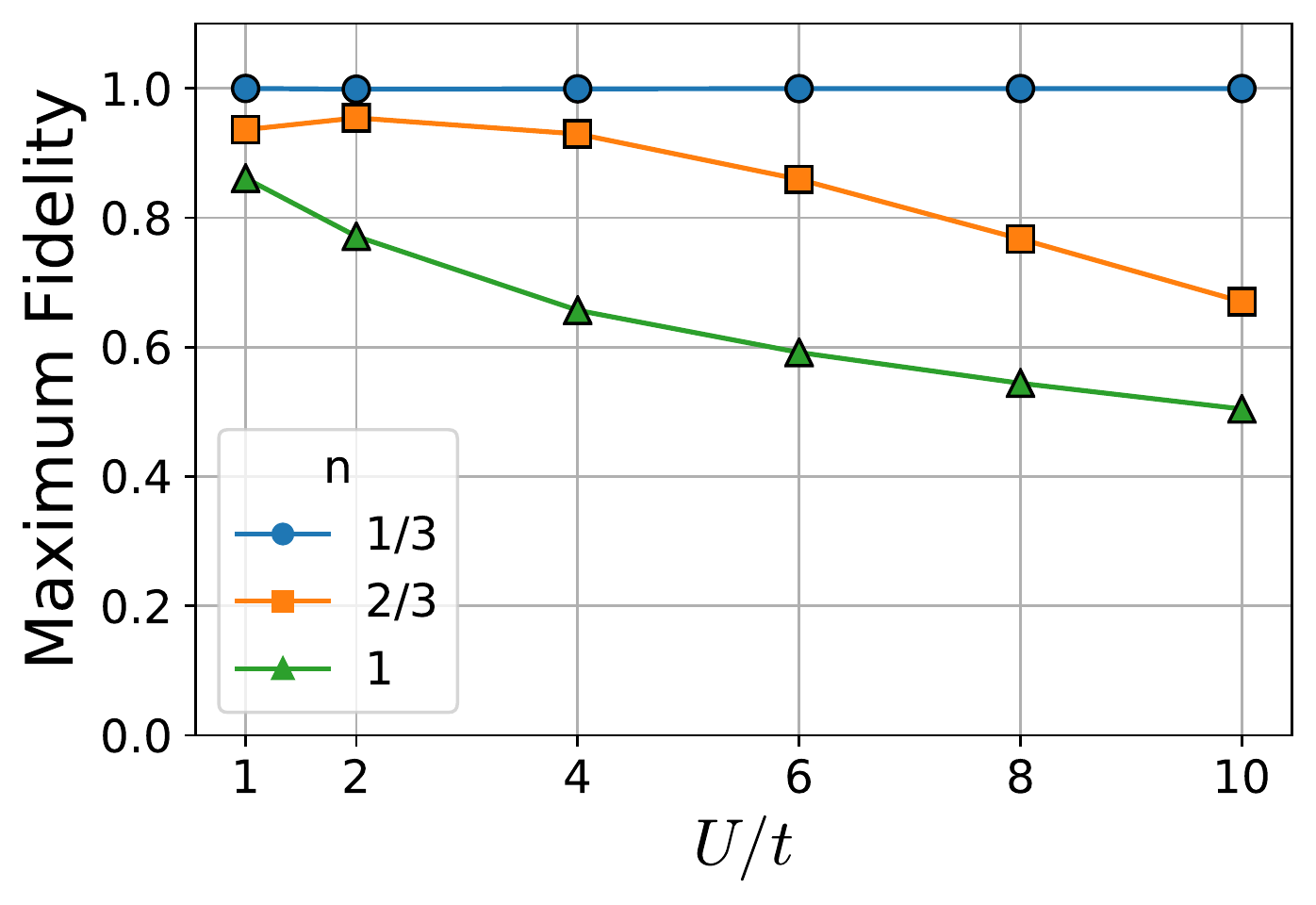}
    \label{fig:hubbard_max_fid_diff}
    }\\
    \caption{Maximum fidelity out of all checkerboard pattern drivers on a $4\times 3$ FHM, between the annealed state and the true $H_{LW}$ ground state at various doping $n$ and interaction strength $U$. All drivers had no terms on the Auxiliary qubits.
    Annealing simulation was performed with a linear schedule and total time $T=100$ in arbitrary units. 
    In (a), the checkerboard pattern for both spins match, where in (b) spin up/down have different drivers. For matching drivers, we show the addition of 2 and 3 $H^{XX}$ Heisenberg driver terms in dotted lines in (a) which greatly improves the fidelity. }
    \label{fig:hubbard_max_fid}
\end{figure}

Next, we use a lattice with two fermionic spins, $\uparrow/\downarrow$, described by the standard Fermi-Hubbard model \cite{Hubbard}. 
\begin{align}
    H_{LW} = -t\sum_{\langle ij\rangle \sigma} c^\dagger_{i\sigma} c_{j\sigma} +h.c. + U \sum_i n_{i\uparrow}n_{i\downarrow},
\end{align}
where $\langle ij\rangle$ denote nearest neighbors with open boundary conditions and fixed $t=1$ (again not to be confused with $t$ used for time in the previous section). We study our method on a $4\times 3$ lattice at various dopings, which is the largest accessible to our annealing simulation using exact simulation methods. Significant classical work has been done around $U/t=8$, where one expects a severe sign problem and evidence of a connection to experimental results, as well as studying intermediate ($\sim 2 \leq U \leq 6$) and strong coupling ($\sim U \geq 6$) regimes \cite{LeBlanc2015,Qin2021}. Thus we simulate a broad range of interaction values $1\leq U \leq 10$ to understand the protocol performance across the phase diagram.

To simulate both fermion spins, we create two sublattices for each spin containing the same 5 non-commuting hopping terms (see Fig.~\ref{fig:model_FH}a), which interact via a density-density term (realized via Pauli $(I-Z)_{i,\uparrow} (I-Z)_{i,\downarrow}$ operators). Due to the increased size of both the Hilbert space and number of possible drivers of the form Eq.~\eqref{eq:driver}, we continue to focus on simulating only checkerboard pattern drivers within each spin subspace. Given the additional Pauli weight of including auxiliary terms in the driver and the strong performance of the spinless models in the previous section, we also focus on drivers without these extra terms. 

Each sublattice now has an independent choice of driver, where both spin up and down drivers are matching or different, or in other words either the initial bulk product state has complete double occupancy (matching) or the initial state has no double occupancy (different). We enumerate the maximum fidelity, squared modulus of the overlap between the ground state and annealed state, for either matching or different drivers in Fig.~\ref{fig:hubbard_max_fid}. Note that although we generally obtain good energetics (see Fig.~\ref{fig:hubbard_min_delE}), we carefully study the fidelity as there is not a large gap in this system and thus the final state may be a mix of low lying states. When the checkerboard pattern for both spins match, we find high fidelity for low $U$ but decaying fidelity as interaction strength increases. Note that despite overall low fidelity, we obtain reasonable energies with a maximum $\Delta E/|E_0|$ of 0.12 and 0.02 for matching and different drivers, respectively.

While energy is an important observable quantity to consider, often one can obtain good energies but lack correct short or long range physics of the true FHM ground state. To ensure not only low energy but the physical properties of the final state, both local and non-local observables such as density $\langle n_i \rangle$, local spin $\langle S^z_i \rangle = \langle n_{i\uparrow} \rangle - \langle n_{i\downarrow} \rangle$, and correlation functions for charge $C(i,j) = \langle n_i n_j \rangle - \langle n_i\rangle \langle n_j\rangle$ and spin $S(i,j) = \langle S^z_i S^z_j \rangle - \langle S^z_i\rangle \langle S^z_j\rangle$ are measured and compared against the true ground state. 

\begin{figure}
    \centering

     \includegraphics[width=\columnwidth]{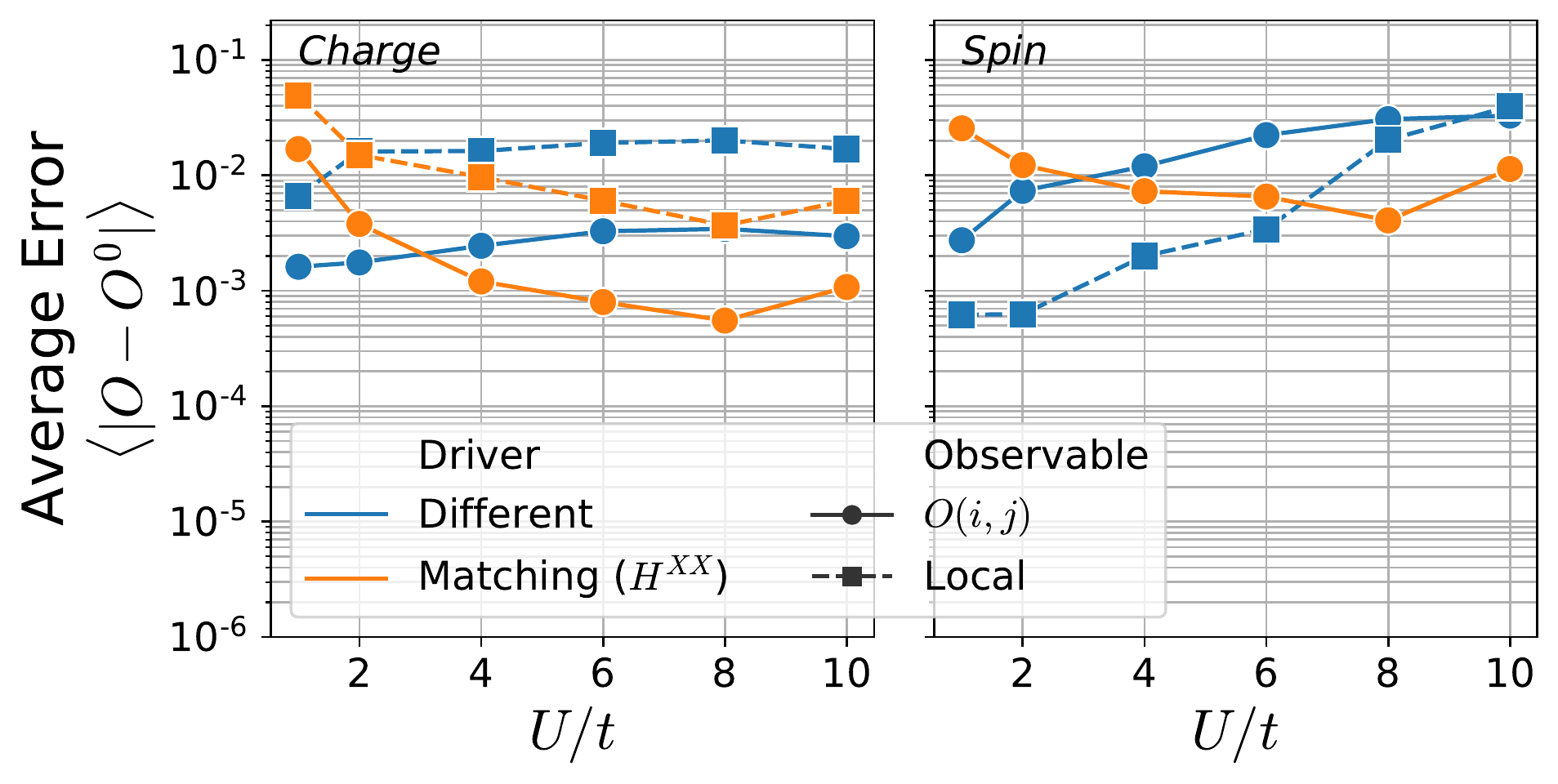}

    \caption{Average observable error over all sites of $n=2/3$ data from Fig.~\ref{fig:hubbard_max_fid} for various values of $U/t$. Charge and spin measurements correspond to left and right plots respectively; blue lines corresponding to different spin drivers and orange corresponding to matching spin drivers with two $H^{XX}$ Heisenberg terms added to improve fidelity, respectively. Note that there is no error for the local spin density on the right plot. }
    \label{fig:hubbard_obs_err}
\end{figure}

We compute the average error $1/N \sum_i |\langle O_i \rangle - O_i^{exact}|$ for spin density $S^z$, charge $n$, and their correlation functions using the checkerboard driver annealed state in Fig.~\ref{fig:hubbard_obs_err}. For both matching and different drivers in each particle number sector the average charge error is very small, at most of order $10^{-2}$. 

\begin{figure*}
    \centering

       \includegraphics[width=.67\columnwidth]{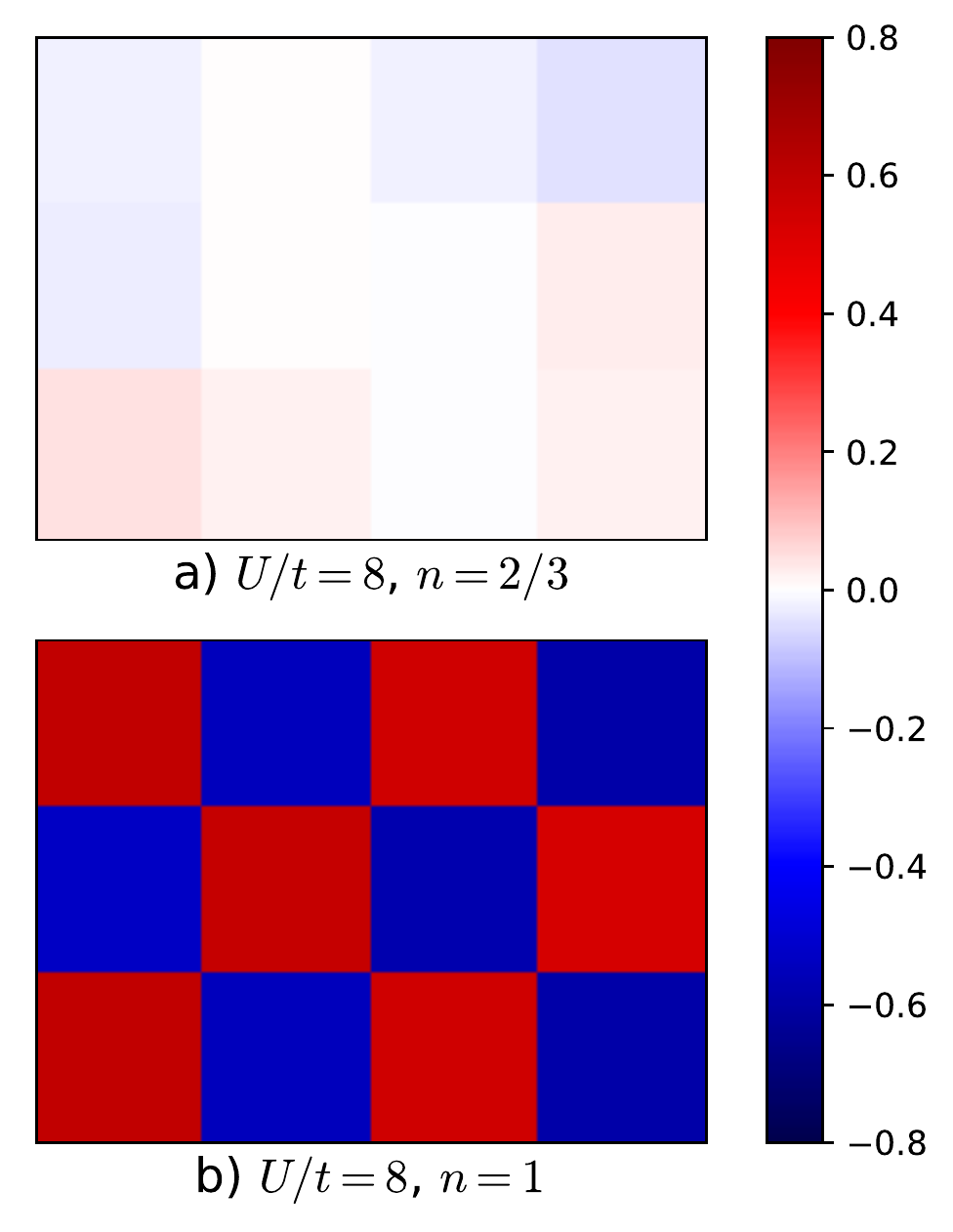}
     \includegraphics[width=.3\columnwidth]{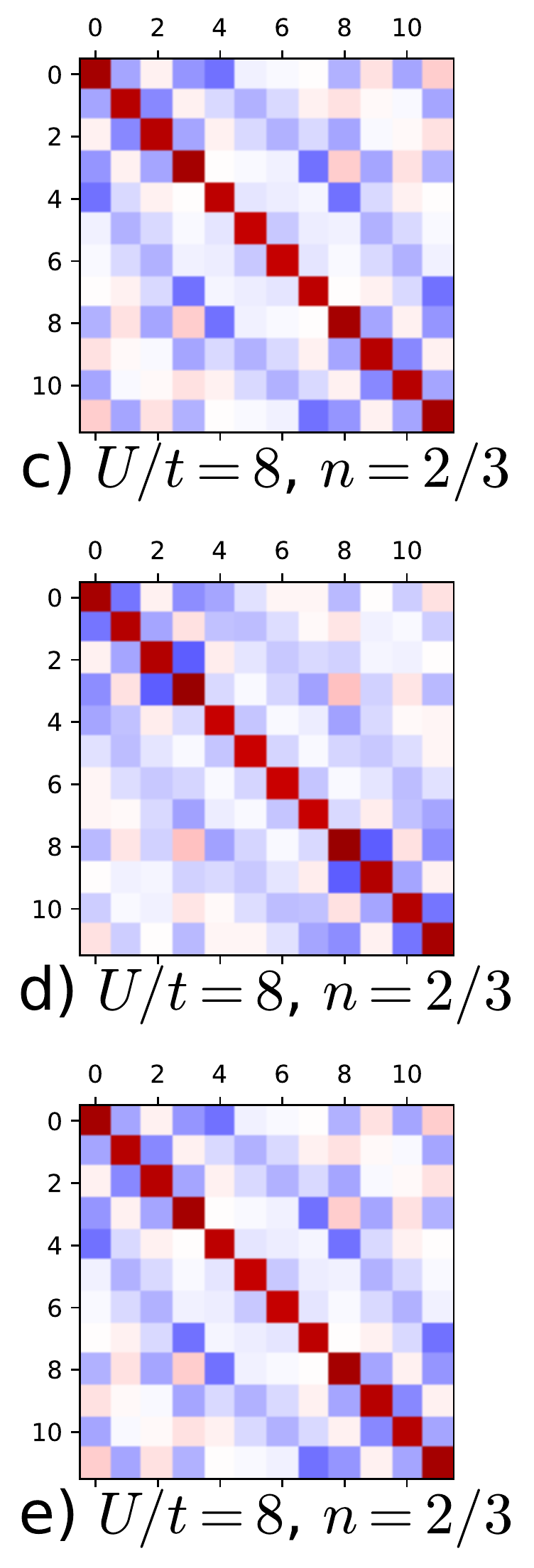}
     \includegraphics[width=.63\columnwidth]{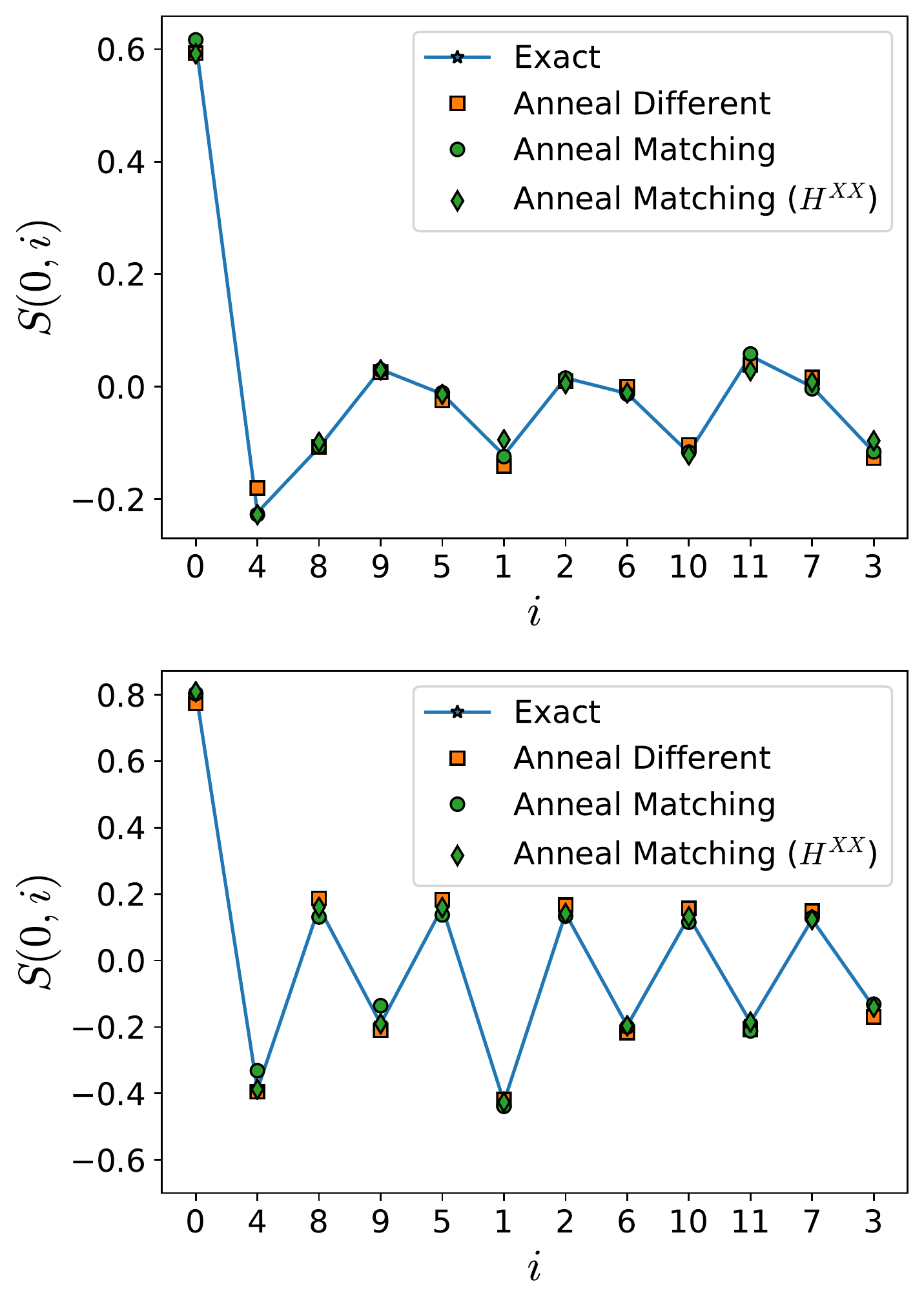}
    \caption{Left: Local spin density $\langle S^z_i\rangle$ of driver with different checkerboard patterns from Fig.~\ref{fig:hubbard_max_fid_same} (a,b) for the $4\times 3$ FHM. The ground state and matching spin drivers generate no local spin density. 
    Middle: Spin-Spin correlation function $S(i,j) = \langle S^z_iS^z_j \rangle - \langle S^z_i\rangle\langle S^z_j\rangle$ for the ground state (c), different spin drivers (d), and matching spin driver with two additional $H^{XX}$ Heisenberg term (e).
t    Right: Spin-Spin correlation function of the first site $S(0,i)$ for $U/t=4$ at $n=2/3$ (top) and $n=1$ (bottom) doping, using the drivers from Fig.~\ref{fig:hubbard_max_fid}.  }
    \label{fig:hubbard_spin}
\end{figure*}

Comparing the local spin density, there is a clear discrepancy between the two driver scenarios. When the checkerboard pattern matches both up and down spin, our driver and Hamiltonian are identical under spin exchange and both the ground state and annealed state have effectively zero ($\leq 10^{-12}$) local spin density despite low fidelity between the annealed state and true ground state. With different checkerboard pattern drivers, there is a strong local antiferromagnetic spin density where swapping the up and down sublattice drivers can produce the reversed pattern, shown in  Fig.~\ref{fig:hubbard_spin}a,\ref{fig:hubbard_spin}b for two different doping. The spin-spin correlations on the other hand, for different or matching spin drivers, shown in Fig.~\ref{fig:hubbard_spin}d,e respectively, retain most of the salient features of the ground state in Fig.~\ref{fig:hubbard_spin}c, but annealing with different spin drivers produces a generally larger correlation than the ground state due strong local spin.  

The mismatch in local spin density and low fidelity as we approach high $U$ is most likely due to the fact that the driver and interaction term $\sum_i n_{i\uparrow}n_{i\downarrow}$ commute. To overcome this with matching drivers, we introduce an additional term to the driver, of a nearest neighbor $H^{XX}$ Heisenberg term $H_{ij}^{XX} = -(X_iX_j+Y_iY_j)$. This term does not commute with the Hubbard $U$ term, but commutes with total particle number; in order to commute with each stabilizer, these additional terms must be restricted to specific locations around the perimeter of the lattice (for lattice dimensions larger than 2). The addition of this term also modifies the initial driver ground state, and increases the maximum Pauli weight to 2. Generation of this initial state, with the addition of an entangled triplet state, can be done via an annealing protocol or thermalization \cite{Li2018}.

In Fig.~\ref{fig:hubbard_max_fid_same} we show the maximum fidelity with the addition of two $H^{XX}$ terms on the upper left/right edge of the lattice in lieu of two $+Z$ terms. There is dramatic improvement in the fidelity, at $U/t=8$ and $n=2/3$ the fidelity goes from $\approx 0.832$ to $\approx 0.948$. This improvement extends for intermediate and large values of $U/t$, but yields a lower fidelity for $U/t<4$. 
% One mitigation of this however, may be adjusting the relative scaling of the XX term and checkerboard pattern as a function of $U/t$.  

While the square FHM at half filling has an efficient classical algorithm \cite{LeBlanc2015}, we include our method's performance for completeness. At half filling, there is strong anti-ferromagnetic correlations in addition to particle-hole symmetry. In order to obtain high fidelity results at large $U/t$, 3 $H^{XX}$ terms must be introduced on the $4\times 3$ lattice ($n_e/2$ $H^{XX}$ terms). The fidelity again increases from 0.324 to 0.981 for $U/t=8$, with a broad improvement across multiple $U/t$ strengths.

We also briefly address the scaling of the gap and the effect of noise on this system.
For the Hubbard model, we lack data across different system sizes. Instead, we can observe how the minimum gap changes as a function of interaction strength $U$. In Fig.~\ref{fig:gap_hubbard} we show the ratio of spectral gap to minimum gap for the $4 \times 3$ lattice as a percentage of the total anneal.  We find that for $n=2/3$, the minimum gap occurs exactly at the end of the anneal at $s=1$ ($t=T$). Similar to the spinless case, the minimum gap is heavily intertwined with the spectral gap which may hint at the worst-case behavior of the protocol to larger systems. 
In Appendix~\ref{sec:app_noise}, we consider noise introduced from control errors, where each of the $H_i$ terms may be implemented with a different coefficient than the intended one. Note that we only consider noise within the logical subspace, such that non-logical final states are still forbidden.  
If the noise is on the order of the spectral gap can maintain similar high fidelity (e.g. from $\approx 0.996$ to on average $\approx 0.988(1)$). These observations however are only meant to point toward viability of this protocol, and further understanding is left to future work.

% \section{Discussion}

\section{Experimental perspectives}\label{sec:experimental}

Our proposal presents a number of features that makes it particularly amenable to experimental realisation. An important selling point is our focus on QA as a platform of choice. Because of lower control requirements, QA has already been implemented and demonstrated at a large scale compared to gate-model quantum computers. Commercial quantum annealers now have qubit numbers exceeding 5000, whereas quantum computer qubit counts have barely exceeded 100 qubits. While the  proposed adiabatic approach cannot be implemented on current commercial quantum annealers, the requirements are within reach for next generations of quantum annealers and are in line with hardware research directions currently being pursued for advanced annealers. A quantum annealer that could carry out our proposed approach could allow early access to solutions of classically intractable FH models in the NISQ era. Because of the close ties between certain gate model and adiabatic quantum approaches \cite{zhou2020quantum,brady2021optimal,brady2021behavior, wurtz2022counterdiabaticity}, exploration of this approach on early advanced quantum annealers would give insight not only into future quantum annealing approaches but also related gate-model approaches. In this section, we outline the resource requirements and experimental prospects for our approach and discuss advantages and limitations compared with other approaches.

\subsection{Architecture choice}

So far the FH models have been realized experimentally on a gate based quantum computer in a variational hybrid quantum-classical algorithm \cite{Linke2018,Montanaro2020,Suchsland2021, Stanisic2021} . The largest model \cite{Stanisic2021} used the JW encoded Hamiltonian, its expectation value as an optimization cost function and a suitably selected circuit Ansatz  to find near the ground state energies. Though other observables displayed larger deviation from the theoretical prediction. The reason for the discrepancy is related to expressibility of the Variational Quantum Eigensolver (VQE) \cite{peruzzo2014variational} Ansatz, that lacks rigorous guarantees of convergence to the ground state. One may expect a trade-off between number of optimization parameters, expressibility and noise contribution, which for near term quantum devices needs to be well-balanced, otherwise it can lead to faulty observations. Furthermore, if one is interested in provable convergence to the ground state, then algorithms based on quantum phase estimation are required. A large body of work has been devoted to solve that problem for the fault-tolerant regime for both {\it ab initio} and time dynamics \cite{kivlichan2020improved,Clinton2020,Babush2018}.

The appropriate resource estimation between various computational paradigms is required to properly assess feasibility of each approach. It is, however, non-trivial task to compare different platforms and algorithms directly, however we will provide a flavor of what one may expect from the near term gate based and annealing simulations. 
% however we will try our best to provide at least the flavor of what one may expect from the near term gate based and annealing simulations.
For illustration, a recent paper by Cai \cite{Cai2020} provided resource estimates for running 5x5 FHM (open boundary conditions) with the VQE Ansatz, and gradient based optimizer. A single iteration of the VQE (including gradient and energy evaluation) is estimated to take $10^5\ s$. In practice Stanisic {et.al} \cite{Stanisic2021} used VQE to converge near the ground state in $\approx5\cdot10^3\ s$  for 2x4 lattice (the largest they simulated). In contrast, annealing energy estimation requires collecting $M\approx 4\cdot10^5$ samples (as suggested in \cite{Cai2020}, however in \cite{Stanisic2021} a varying number of shots is used for optimization (1000) and final estimation (100,000)) for each of the five measurements, thus we get the total time estimate
\begin{equation}
    t_{\mathrm{TOT}} = 5M (t_a + t_r + t_d),
    %\approx 5\cdot10^4 (100 + 50 + 21)\mu s = 6.55 s,
\end{equation}
where $t_a$ is annealing time, $t_r$ is readout time and $t_d$ is QPU delay time. Using data for the D-Wave Advantage hardware \cite{DWaveAdvantage}: $t_a\in[0.5,2000]\ \mu s$, $t_r\in[25,129]\ \mu s$ and $t_d=21\ \mu s$, which brings us to the total time estimate $t_{\mathrm{TOT}}\approx [10^2,5\cdot10^4]$. It is clear, that based on these approximate QPU estimates, the annealing protocol is an attractive competitor for NISQ era gate based algorithms, falling between numbers reported in \cite{Cai2020} and \cite{Stanisic2021}. However, one needs to be careful in making this direct comparisons, since two platforms runs not only two different algorithms, but also operate under different paradigms. The approach presented in \cite{Stanisic2021,Cai2020} also exploit some error mitigation strategies to reduce noise impact. Additionally, the VQE Ansatz from \cite{Cai2020} with 25 layers gives no guarantees on the convergence and accuracy of the algorithm, as well as it is unknown how many iteration would be required to achieve satisfactory accuracy. In the case of experimental realization for 2x4 lattice, Stanisic {et. al} in Ref.~\cite{Stanisic2021} used a single layer VQE  with only 4 variational parameters. Nevertheless they were  still able to reach an absolute energy error of e.g. $10^{-2}$ for $U/t=4$ at half filling. However, it is unclear how their Ansatz scales, and how accurate it can be for larger systems. On the other hand, in this contribution we have not explored detrimental effects of noise, in particular how it can affect leakage to non-logical subspace, and what methods can be employed to mitigate this effect. Also, in these estimates we assume access to the required initial state, which may require additional overhead (e.g. related to thermalization or annealing state preparation) that is longer than one reported in $t_d$ of D-Wave. Additionally, in \cite{Cai2020} the JW transform is used, while our approach uses the LW encoding that operates on more qubits. Last, but not least, the annealing protocol is size-independent\footnote{Clearly one requires more sophisticated hardware to control the evolution, and one expects a decreasing gap for larger systems, leading to longer-annealing times. However, these are the same problems one would need to face in the gate model, which on top requires deeper circuits to properly mediate important interactions - i.e. deeper layers in VQE.} in terms of running times, and the accuracy seems to be fairly stable across the numerically investigated systems. 

There exist alternatives to VQE in a gate based model, namely those via a Hamiltonian evolution model (see e.g. Ref.~\cite{dong2022ground} for a comparison of different approaches). Many such approaches require many multi-control qubit gates or at the minimum query access to $\exp[-iH]$. On near-term devices, this is often achieved via a trotter decomposition; thus while a potentially attractive method in a gate based model, the ability to fully implement the LW encoded Hamiltonian directly in annealing hardware can be see as advantageous compared to the requirements to implement Hamiltonian evolution. 
We leave more direct comparison for future work. Nonetheless, these rough estimates demonstrate that the annealing proposal can be an interesting alternative for solving FHM on the near term hardware.     
%to experimentally demonstrate the solution of a $4\times 3$ lattice, our method would require [qubits/couplers], compared to such and such in a gate-model implementation. \rl{How accurate should it be? Can't reliably compare quantum simulation implementations. Quantum annealing encoding exact hamiltonians is not an aim currently. Open questions on hamiltonian emulation. These kinds of emulators allow much stronger coupling compared to the gate model, much more nonlinear QM, more non-commuting terms. Gate model introduces trotter errors when using strong coupling terms. Value in trying drastically different approaches to solving FH.}

\subsection{Qubits, couplers and measurements}

The LW fermion-to-qubit encoding presents a possible advantage for implementation of FH problems on superconducting QA hardware. In our proposal, the interaction order is restricted to three. The experimental implementation of local X- and Z-fields and two-local ZZ interactions \cite{Maassen_van_den_Brink2005,Harris2007,Harris2009,Weber2017} has been established for a long time. However, the need to implement interactions of order greater than two has been of great interest for constructing universal QA \cite{Albash2018,Aharonov2008} and enabling more widespread applications such as the simulation of highly correlated fermionic systems \cite{Albash2018,Bravyi2008}. The earliest proposals for this are based on perturbative gadgets, which provide a way to build arbitrary $k$-local interactions from two-body interactions \cite{Jordan2008}. These proposals require a large number of auxiliary qubits and two-local couplers and further they do not provide a clear route to the implementation of physical circuits. 
On the other hand, since then we have also witnessed ideas to implement directly (i.e. without perturbative techniques) interaction terms of order three or more on the real superconducting platform  \cite{Leib2016,Chancellor2017}. %% TOFIX: What is real?
% There have since been proposals to implement real superconducting circuits that provide interaction terms of order three or more without using perturbation theory \cite{Leib2016,Chancellor2017}.
These approaches, however, only approximate the ground-state and also require auxiliary qubits. More recently, native interactions of order three \cite{Melanson2019} and four \cite{Schondorf2019} have been studied, their realization relies on a tailored nonlinearity, rather than through emulation via a set of subsystems.
% More recently there have been proposals to natively create interactions of order three \cite{Melanson2019} and four \cite{Schondorf2019} through a tailored nonlinearity, rather than through emulation via a set of subsystems. 
Furthermore, computational methods for finding the best circuit configuration given a set of requirements have been demonstrated and applied for a four-local ZZZZ coupler design that is robust against noise and tailored for flux qubits \cite{Menke2021}. Menke {\it et al.} \cite{Kerman2018aps1, Menke2021aps1, Menke2021aps2, Menke2022} are actively working on demonstrating a four-local coupler experimentally.
%There is ongoing work to demonstrate experimentally a four-local coupler[T. Menke]. %Although there are not yet any published experimental results, experiments on a four-local coupler have been performed and show that in principle they can be implemented[Tim Menke].

Experimental high-order coupler proposals have explicitly concerned the implementation of ZZZ \cite{Melanson2019,Menke2022} and ZZZZ \cite{Schondorf2019, Menke2021} interactions, as these are arguably the most relevant to quantum combinatorial optimization annealing protocols, in which the final problem Hamiltonian is diagonal. However they need not be restricted to the Z basis. A recent proposal for a novel qubit design, the Josephson phase-slip qubit (JPSQ), could be used to fully emulate the spin degrees of freedom of a spin-$\frac{1}{2}$ particle \cite{Kerman2019}. This allows the X, Y and Z degrees of freedom of the qubit to be independently controlled. Crucially the JPSQ (see Section 4 and Fig. 12 in \cite{Kerman2019}) allows the coupling of arbitrary degrees of freedom between qubits, which is of great interest in the quantum annealing community as a whole. With this qubit it is within realms of feasibility to implement complex interaction terms such as XXY by inductively coupling the relevant loops of the qubit with the multi-body coupler proposals discussed in the previous paragraph. There is ongoing experimental work to fabricate and measure JPSQ devices \cite{Hirjibehedin2020aps1, KermanFuture}. Achieving that goal, would bring us closer to a platform capable of simulating the discussed FHM annealing protocol

We require a measurement protocol that at a minimum yields the energy expectation of the FHM Hamiltonian $\langle H_{LW} \rangle$. The expectation values of each of the terms in the Hamiltonian, as described by Eqs. (6)-(8), can be used to construct the energy expectation value of the FH Hamiltonian. There is currently no established method of directly measuring the expectation value of non-local observables such as ZZ or XXY. They can however be estimated from repeated measurements in each of the local X, Y and Z bases. A powerful readout technique for performing measurements in a specific basis, called the persistent current readout, has been demonstrated experimentally for high-coherence qubits \cite{Novikov2019, Schondorf2020, Grover2020} and has also been used by D-Wave for more than a decade \cite{Berkley2009}. This readout technique presents significant advantages over previous dispersive techniques \cite{Lupascu2006}, including very high fidelity and single-shot capabilities. Most importantly however, since the JPSQ can be designed such that persistent currents reflect the spin of a qubit in all degrees of freedom, this readout technique can be used to measure the JPSQ state in the X, Y and Z basis.

\section{Conclusions/Outlook}
\label{sec:conclusions}

% Summary of the results and their potential implication on constructing an adiabatic device. Discuss problems 
% \begin{itemize}
%     \item scaling - we can't guarantee performance for large systems, though simulations indicate some hope
%     \item errors/noise - briefly discuss need for understanding what can happen if we start include realistic noise models (pose this as an open problem), and maybe talk about need for error correction schemes here, that might be achievable easier (obviously subjected to a noise model) than for a universal QC, since we have already a favourable geometry,
%     \item different schedules/drivers and non-adabatic models
% \end{itemize}
% \fifix{We may consider rephrasing the conclusions to put stress on different things. Maybe starting with (numerical) investigation of annealing protocol under LW, and later potential realization...}
We proposed and numerically investigated a maximum Pauli weight 3 scheme using (coherent) quantum annealing to solve a $t-V$ spinless fermion model and the Fermi-Hubbard model on a square lattice. Several driver solutions were developed around using $\pm Z$ terms in a checkerboard pattern: a completely local driver which has good fidelity, local observables, and correlation functions for small and intermediate $U/t$. The same driver with the addition of a number of $H^{XX}$ terms along the perimeter to maintain high fidelity in the intermediate and large $U/t$ regime. 

% Using a driver based on $\pm Z$ terms in a checkerboard pattern, we obtain high fidelity results across a variety of lattice sizes and dopings. In the intermediate to large $U$ regime for the FHM we present a solution for obtaining high fidelity results using  an additional XX Heisenberg like term and ensure the lack of local symmetry breaking.

% \fifix{This requires more substance}
While the focus thus far has been on the LW encoding, our results extend generically to fermionic encodings which have efficient initial state preparation. Each encoding can be viewed as a unitary rotation of the JW encoding (if there are any additional phases introduced compared to the Jordan-Wigner encoding, like in the LW encoding), and as such the energy results from our protocol will be identical (provided one operates in the coherent evolution framework). For encodings that have a similar stabilizer or gauge condition structure, the product state driver can also be written naturally in those encodings, see for example the method outlined in \cite{Whitfield2016}. Our developed protocol can then be applied to a number of alternative methods, if a different topology or encoding of operators becomes feasible for future hardware. 

There are a number of further questions left for future work. For instance, how to robustly show the scaling of quantum annealing of the FHM. Despite the lack of guarantee, our results have hopeful indications. At low doping we can scale to large lattices of $6 \times 6$, which did not hinder the final fidelity. Our protocol also provides a polynomial sized set of potential drivers, such that how the protocol scales is clear.  

% The second question, is the role of noise in simulations. The underlying encoding still retains error correction properties and thus could be utilized both during the annealing process and in the measurement scheme \rl{cite}. Given the favorable geometry in this method, error correction schemes could be easier to implement than a general purpose quantum computer. Beyond the encoding itself, our method removed Pauli weight 4 terms of the driver Hamiltonian in order to produce a fully local (with a potential XX Heisenberg term) driver. If a given noise model has the ability to include non-logical states during the annealing process, driver terms which are localized to the auxiliary qubits may become more relevant. 

The second question, is the role of noise, that one may expect in the real hardware realization. %Moreover, three-way couplers necessary to implement Pauli weight 3 terms would require higher decree of control than lower weight terms.  
In the LW encoding annealing approach, noise may be more detrimental than in the case of standard annealing setups offered by D-Wave. LW encoding introduces notion of logical space, and the true physical (operating on all data and auxiliary qubits) ground state lays lower in the energy spectrum than the logical ground state. Hence, any leakage to the non-logical state deteriorates the observed results, in particular one may expect natural relaxation and thermalization processes that could lead to leakage. This, in principle, can be modeled with adiabatic master equation \cite{Albash_2012}, however access to large system sizes in that case is more restricted by the density matrix simulation. The underlying encoding still retains error correction properties and thus could be utilized both during the annealing process and in the measurement scheme \rl{cite}. Given the favorable geometry in this method, one may expect to devise and implement the error correction schemes similar to \cite{Mohseni_2021}, easier than in a general purpose quantum computer, though this problem is non-trivial and we leave it for future contribution. Beyond the encoding itself, our method removed Pauli weight 4 terms of the driver Hamiltonian in order to produce a fully local (with a potential $H^{XX}$ Heisenberg term) driver. If a given noise model has the ability to include non-logical states during the annealing process, driver terms which are localized to the auxiliary qubits may become more relevant. Another possibility to circumvent noise problem is to include penalty terms, as proposed in \cite{Pudenz_2014} based on the stabilizer formalism, this however would require Pauli weight 8 terms, which is beyond the current proposal. 

Finally, the question of schedule optimality remains open. We presented results only using a linear schedule, but we find that our results are relatively robust for other schedules with the same total time, namely those similar in use on D-Wave systems (see Appendix~\ref{sec:app_schedules}). Hardware limits may encourage shorter total annealing times as well as schedule considerations not undertaken in our simulations. % and the question of whether an overall optimal schedule exists remains open 
Relaxing the condition of adiabaticity may also lead to improved results on future hardware \cite{Crosson2021, FryBouriaux2021, Khezri2022}, in particular introduction of counterdiabatic form might be a promising avenue for improving the protocol \cite{XieCounter}.

% \fifi{Even though the currently available hardware is still behind this proposal, some promising perspectives are suggesting that the introduced protocol can be experimentally realized in the near future. Therefore, placing this application specific quantum annealing hardware on a map of potential candidates to achieve quantum advantage. Through our numerical and theoretical analysis we have shown that the annealing protocol is an attractive alternative for gate based algorithms, at least if considering non fault-tolerant regime. }
Even though there is no currently suitable quantum annealing hardware to run our introduced protocol, some promising perspectives suggest that the protocol can be experimentally realized in the near future. 
Therefore, placing this application specific quantum annealing hardware on a map of potential candidates to achieve quantum advantage. Through the lens of a non-fault tolerant regime, we have shown via numerical and theoretical analysis that the annealing protocol is an attractive alternative for gate based algorithms.

\section*{Acknowledgements}
Authors would like to thank A. J. Kerman, R. LaRose, L. Brady and P. A. Lott for insightful discussions and comments. 
We are grateful for support from the DARPA RQMLS program under IAA 8839, Annex 128 and from the U.S. Department of Energy, Office of Science, National Quantum Information Science Research Centers, Co-design Center for Quantum Advantage (C2QA) under contract number DE- SC0012704 including through NASA-DOE interagency agreement SAA2-403601.
The work supported by C2QA was primarily the resource analysis (gap scaling, noise modeling, and the gate-annealing comparison), with the DARPA RQMLS program supporting the model construction, annealing evolution, and annealing experimental perspectives. 
RL and JB are supported by the USRA Feynman Quantum Academy funded by the NAMS R$\&$D Student Program.
FW, ZGI, ZW, and JM are thankful for support from NASA Academic Mission Services, Contract No. NNA16BD14C. LF-B and PAW are supported by EPSRC grant ref. EP/T001062/1. PAW is additonally supported by EPSRC grant refs. EP/W00772X/2 and EP/W027003/1. DTO'C is supported by an EPSRC-BT CASE award studentship.    

\bibliographystyle{quantum}
\bibliography{refs}
\clearpage

\appendix
\onecolumngrid
% \section*{Appendix}
\renewcommand\thefigure{A\arabic{figure}}  
\renewcommand\thetable{A\arabic{table}}  

\section{LW Encoding and Driver Example}\label{sec:app_example}
\begin{figure}
    \centering
    %\includegraphics[width=0.5\columnwidth]{Appendix/3x2.png}
    %\caption{Site ordering of the $3 \times 2 $ lattice. }
    \subfloat[]{\includegraphics[width=0.25\textwidth]{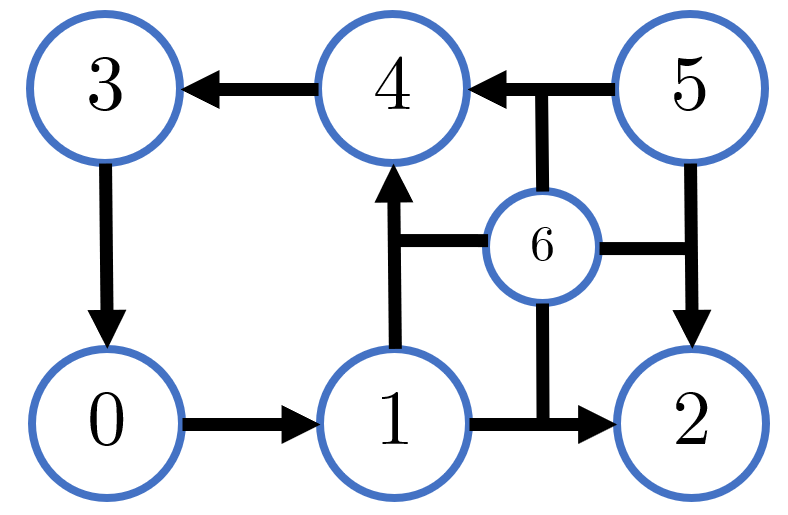}}
    \subfloat[]{\includegraphics[width=0.25\textwidth]{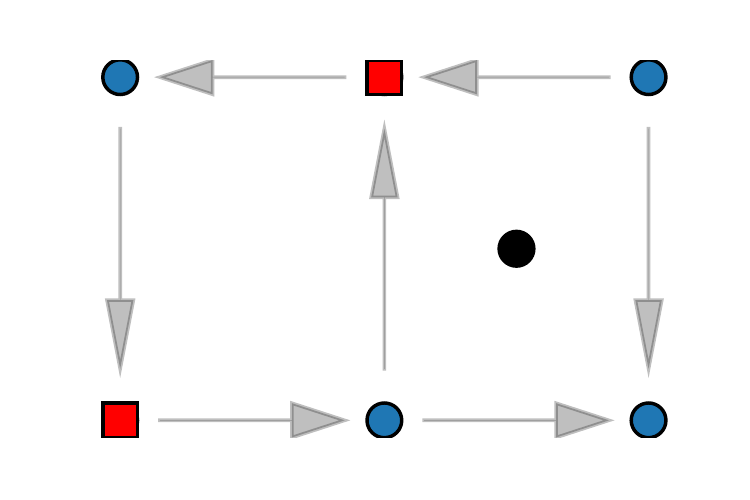}\label{fig:app_3x2_checkerboard}}
    \caption{a) Site ordering of the $3 \times 2 $ lattice. b) Checkerboard pattern driver for $3\times 2$ lattice, with ``on'' bits shown in red squares. The initial product state in the bulk would be $c^\dagger_4c^\dagger_0 |0\rangle^{\otimes 6}$.}
    \label{fig:app_3x2}
\end{figure}
We present the encoded Hamiltonian for two faces, i.e. a $3 \times 2 $ lattice of spinless fermions with site ordering shown in Fig.~\ref{fig:app_3x2}. In general, we order the lattice starting from the bottom left row by row from 0 to $N-1$, shown in Fig.~\ref{fig:app_3x2}a. The hopping Hamiltonian $H_t = -t\sum_{\langle ij\rangle} c^\dagger_ic_j + h.c. $ is encoded as $H_t = -t/2 H_{\text{hor}} + t/2 H_{\text{vert}}^+ - t/2 H_{\text{vert}}^-$ given by 
\begin{align*}
    H_{\text{hor}} =& X_0 X_1+Y_0 Y_1 
                     +X_1 X_2 Y_6 + Y_1 Y_2 Y_6 \\
                    &+ X_5 X_4 Y_6 + Y_5 Y_4 Y_6  + X_4 X_3 + Y_4 Y_3\\
     H_{\text{vert}}^+ =&  X_4 X_1 X_6 + Y_4 Y_1 X_6 \\
     H_{\text{vert}}^- =&X_0 X_3 + Y_0 Y_3 
                        +X_2 X_5 X_6 + Y_2 Y_5 Y_6.
\end{align*}
The interaction term $H_V =  V\sum_{\langle ij\rangle} n_i n_j,$ is similarly encoded  into
\begin{align*}
    H_V = \frac{V}{4}\sum_{\langle ij \rangle } (I-Z)_i  (I-Z)_j.
\end{align*}

% \begin{figure}
%     \centering
%     \includegraphics[width=0.5\columnwidth]{Appendix/3x2_checkerboard.pdf}
%     \caption{  }
%     \label{fig:app_3x2_checkerboard}
% \end{figure}
We can form a checkerboard pattern for two spinless fermions, one of which is shown in Fig.~\ref{fig:app_3x2_checkerboard}. Due to symmetry, this is only one unique checkerboard pattern for this lattice, which has a final annealing fidelity of $\approx 0.9997$. The driver for this pattern is explicitly 
\begin{align*}
    H_D = \textcolor{red}{+}Z_0 - Z_1 - Z_2 -Z_3 \textcolor{red}{+} Z_4 - Z_5.
\end{align*}

Because this driver is only defined on the bulk qubits, we must specify the initial auxiliary state such that the initial state is logical. At $t=0$ of the annealing, the initial state would be $c^\dagger_4c^\dagger_0 |0\rangle^{\otimes 6} \otimes |+\rangle $ where $|+\rangle$ is the +1 eigenstate of Pauli $X$. 

We can additionally add driver terms on the auxiliary lattice, such that the ground state of the driver is exactly the initial state as well as gapped. This is equivalent to an additional $-X$ term to ensure the initial driver ground state (i.e. product state) is logical. 
Adding this term introduces another unique checkerboard pattern driver Hamiltonian, but both have $\approx 0.999$ fidelity. 

\section{Initial State Preparation}\label{sec:app_stateprep}
To ensure the ground state of the driver, here assumed to be composed of $\pm Z$ terms (e.g. checkerboard pattern driver), is a logical state, the auxiliary qubits must be prepared in a specific state akin to a toric code groundstate. This restriction is the same as measuring all the stabilizers to be in the $+1$ eigenstate for a given product state in the bulk. We remark on several possible ways to prepare this state.

There exists an optimally efficient protocol for gate based computers \cite{Higgott2021,Satzinger2021}, which can be summarized as running a circuit of Hadamard and controlled-not gates. One could then take the resulting state prepared on a gate based computer and utilize it in our annealing protocol. The reduced gate set for the gate based preparation may also lend to high fidelity results on near term hardware, and has already been experimentally realized (e.g. \cite{Satzinger2021}).

In addition the gate based routine, we can suggest alternative methods that may align better to native annealing based hardware.  Using the inverse methods of ref.~\cite{Chertkov2018}, one could construct a custom Hamiltonian where the initial auxiliary state was an eigenstate and performing an initial quench or annealing protocol to produce the state. However, there is not a guarantee that this eigenstate will exist with the topology/gates of the annealing device nor the position in the spectrum of the Hamiltonian. Given the number of effective ancilla (the bulk qubits connected to the auxiliary qubits) and the existence of a Pauli weight 2 Hamitlonian which realizes the appropriate state at third order \cite{Brell2011}, this idea may be useful to revisit once specific device capabilities are clear. 

Finally, using measurements alongside a post-processing method is another avenue to prepare the state. In the field of error correction codes, there are methods of state preparation involving rounds of measurements using ancilla qubits \cite{pachos2012introduction}. Bulk qubits could be used in lieu of an additional auxiliary qubit for measurements. Additional protocols may exist that take advantage of this additional qubit structure, which we leave for future work. 

To sum up, our proposal operates under the assumption of having access to the initial state, which has unspecified preparation protocol that would reflect annealing hardware constraints. However, as described above a number of various techniques exists that can inspire hardware efficient protocols. In this case, we identify hardware efficient, as having access to the same device components as the ones required for the dynamics, i.e. couplers of order at most three (connection of two bulk qubits and one auxiliary), and no direct connectivity between auxiliary components (if one has that access in form of couplers of order four, one could readily use adiabatic state preparation protocol discussed in \cite{Hamma_2008}). These constraints could support techniques based on thermal \cite{Babbush_2013}, or gadget-based annealing state preparations. We leave more in depth analysis of hardware efficient state preparation protocol as an open problem.

% Another method could be the use of ancilla based state preparation 
% [Alternative, use of ancillas to measure collapse into the correct state]
% prob success 
%  \rl{experimental perspective here?}

\section{Robustness to Different Schedules}\label{sec:app_schedules}
\begin{figure}
    \centering
    \subfloat[]{\includegraphics[width=0.5\textwidth]{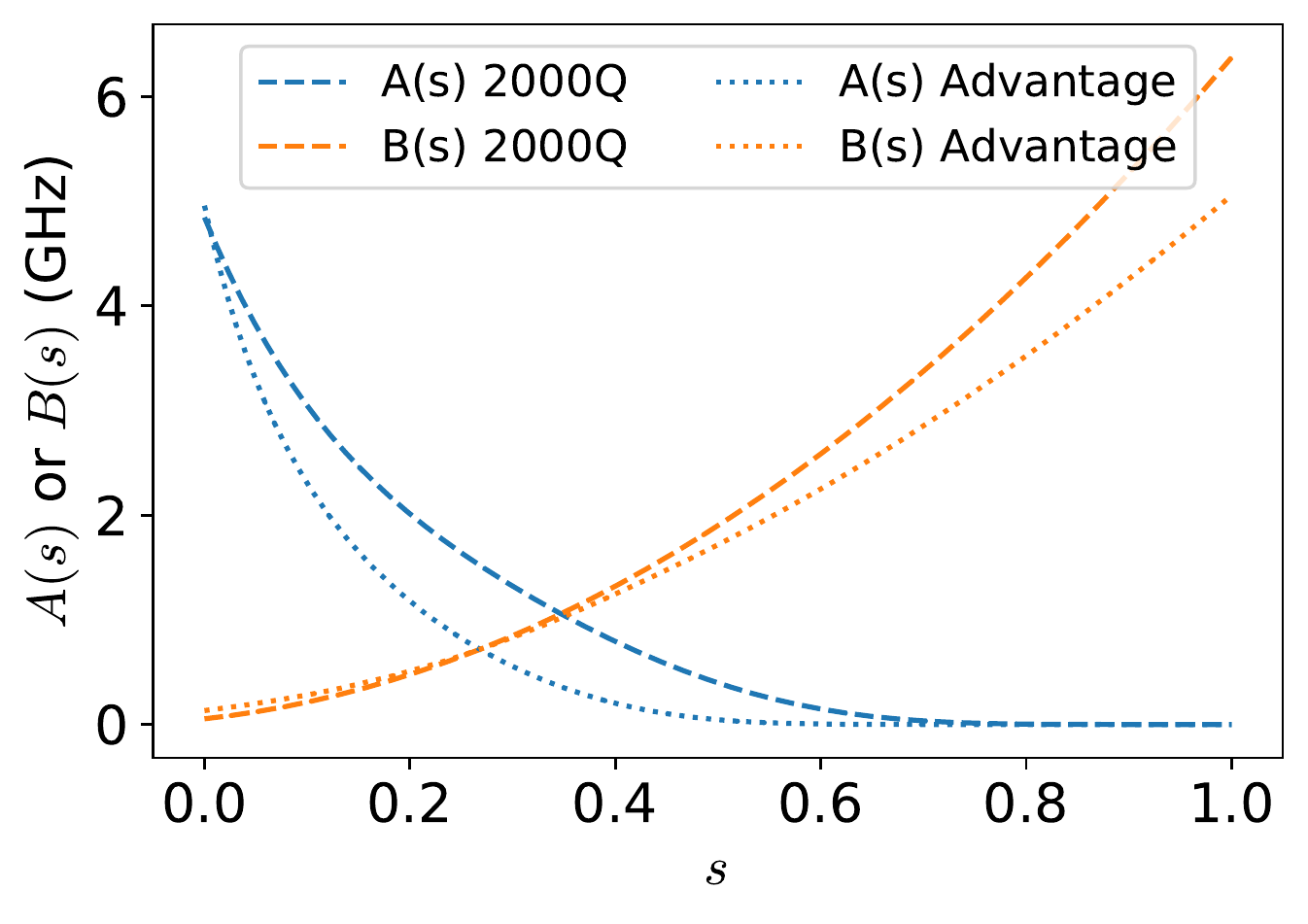}\label{fig:app_schedules_sched}}
    \subfloat[]{\includegraphics[width=0.5\textwidth]{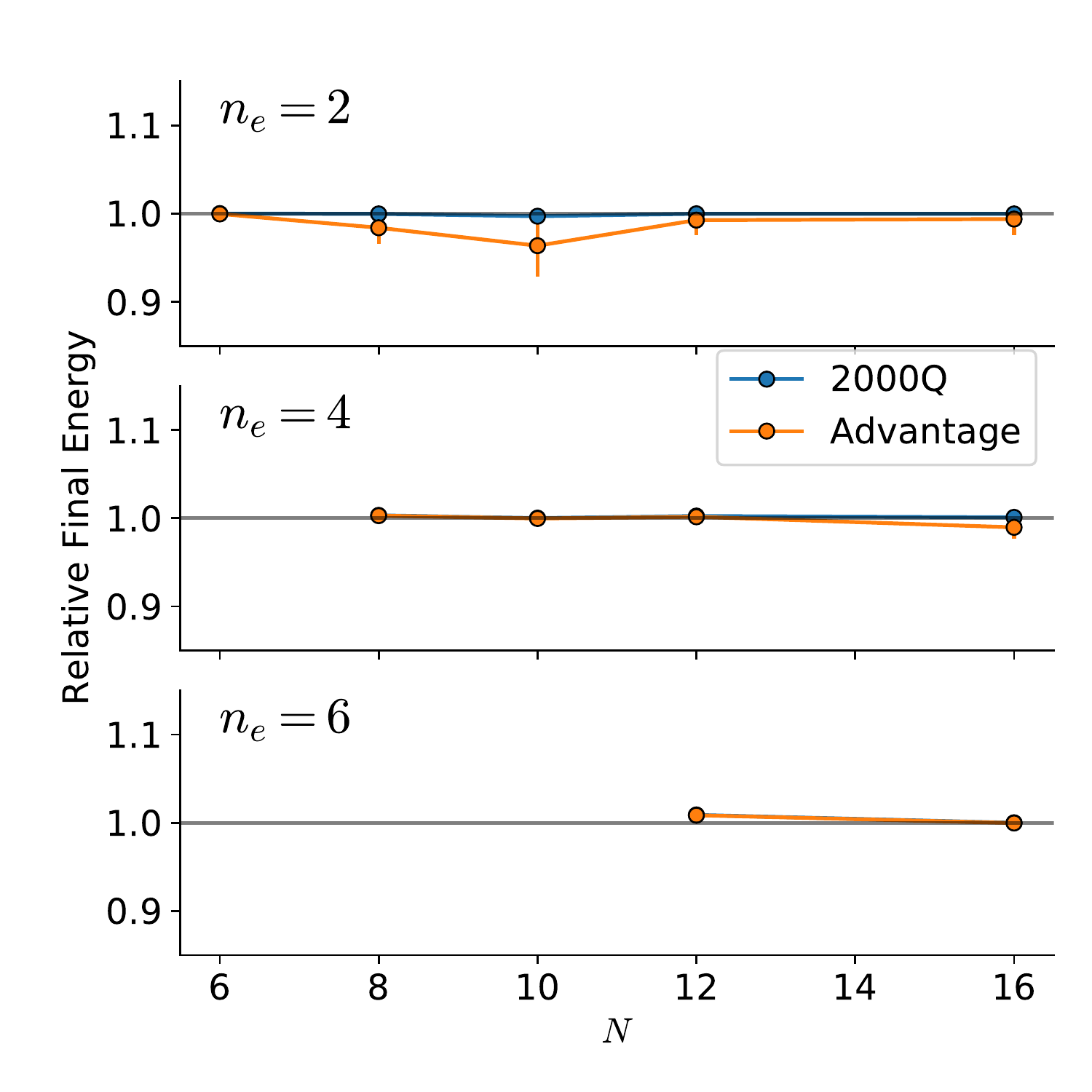}\label{fig:app_schedules_ratio}}
    \caption{a) Schedule functions for the 2000Q and Advantage systems, in units of GHz. Note we have absorbed the factor of 1/2 into these schedules rather than in eq.~\ref{eq:annealing}. b) Relative energy $E^{sched}/E^{linear}$ of spinless fermion annealing at $V/t=1$ with checkerboard drivers. Annealing simulation was done with total time $T=100$ for both schedules. Error bars represent the range over all possible checkerboard drivers.   }
    \label{fig:app_schedules}
\end{figure}

We repeat several of our spinless benchmarks with non-linear schedule functions with the same total time, which we call Sched1 and Sched2. These schedules are based off of schedules from the those implemented in the D-Wave 2000Q \cite{DWave2000Q} and D-Wave advantage \cite{DWaveAdvantage} respectively, shown in of Fig.~\ref{fig:app_schedules_sched}. In Fig.~\ref{fig:app_schedules_ratio}, we simulate the ratio of $E^{sched}/E^{linear}$, where a ratio above 1 corresponds to a non-linear schedule obtaining a lower final energy than the linear schedule.  We find approximately the same energy in nearly all cases, with a small decrease in performance only for occasional system sizes.

\section{Two Particle Correlation} \label{sec:app_tw_corr}
\begin{figure*}
    \centering
    \includegraphics[width=\columnwidth]{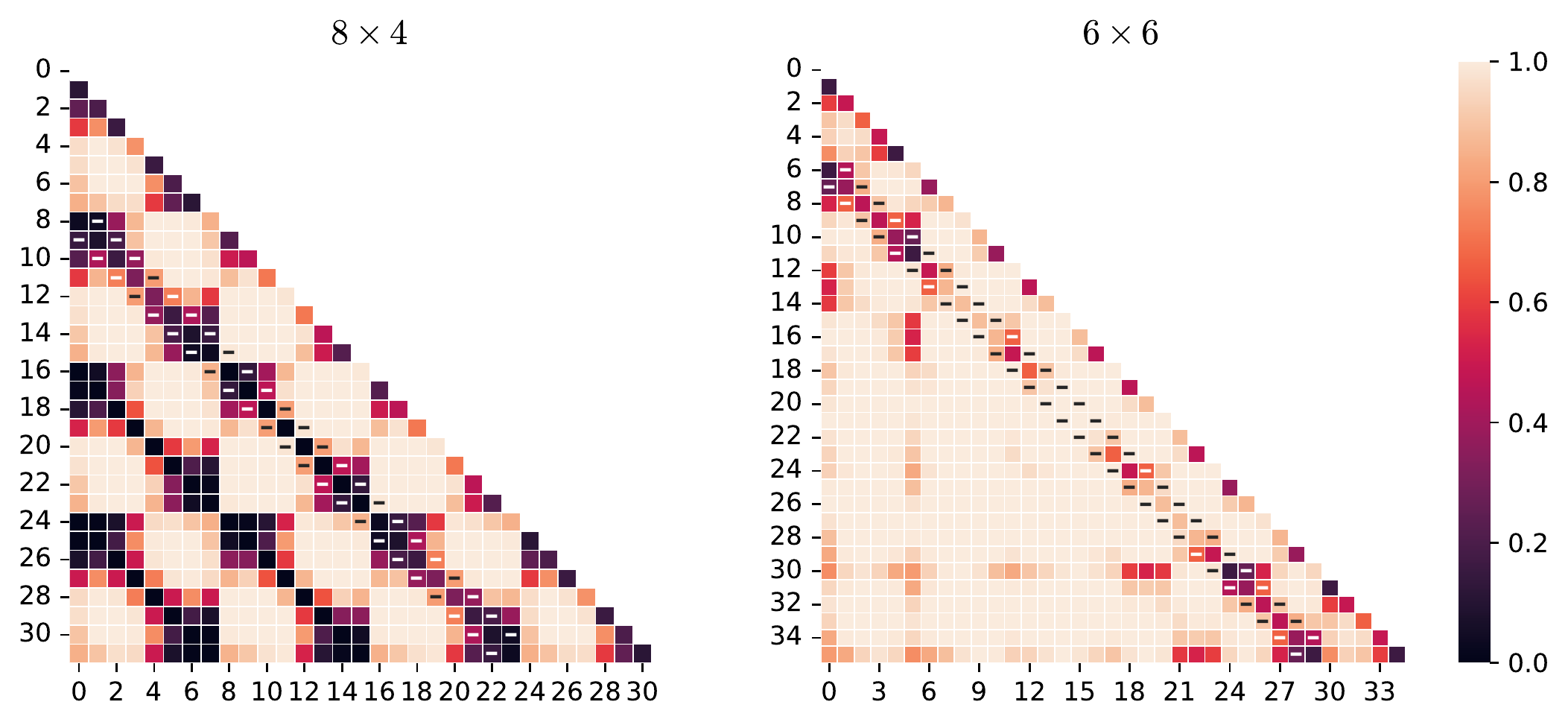}
    \caption{Fidelity for a $8\times 4$ (left) and $6 \times 6$ (right) spinless fermion model at $V/t=1$ using the driver in eq.~\ref{eq:driver}. Axes depict the location for the two fermions (see Fig.~\ref{fig:model_FH} for site ordering). Dashes mark locations of the checkerboard drivers. }
    \label{fig:app_ne2_corr}
\end{figure*}

Using the entire set of $n_e=2$ drivers for a spinless fermion $t-V$ model in the form of eq.~\ref{eq:driver}, we can understand if there are certain locations for the initial state that have low fidelity. We illustrate two examples in Fig.~\ref{fig:app_ne2_corr} for a ladder like lattice ($8\times 4$) and a square $6 \times 6$ lattice. In both lattices, when placing one initial fermion in the corner, overall there is reduced performance compared to other locations. This is particularly clear in the square lattice where the corners (0, 5, 30, and 35) show dark bands (indicating poor fidelity) for most spots on the lattice. 
In general, the best fermion locations for these $n_e=2$ drivers appear to be in the bulk of lattice rather than near the edges, although many exceptions exist. As a result of these correlations, we exclude the patterns that include corners of the lattice in our analysis.  

% explicit n_e2 checkerboard is special case

\section{Fermi-Hubbard Model Energies}\label{sec:app_energy}
\begin{figure}
    \centering
    \subfloat[Matching spin drivers]{
       \includegraphics[width=0.5\columnwidth]{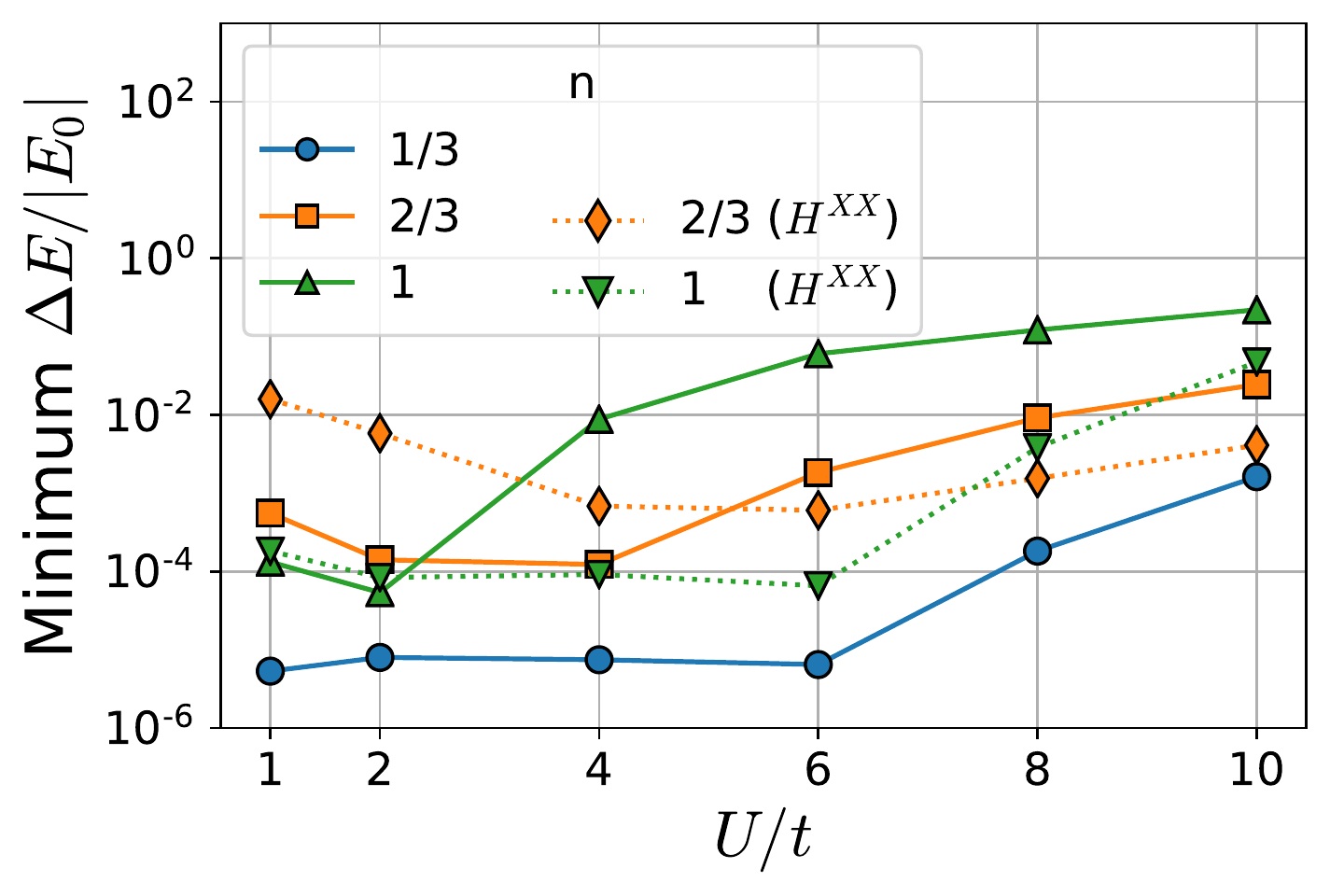}
    \label{fig:hubbard_min_delE_same}
    }
        \subfloat[Different spin drivers]{
       \includegraphics[width=0.5\columnwidth]{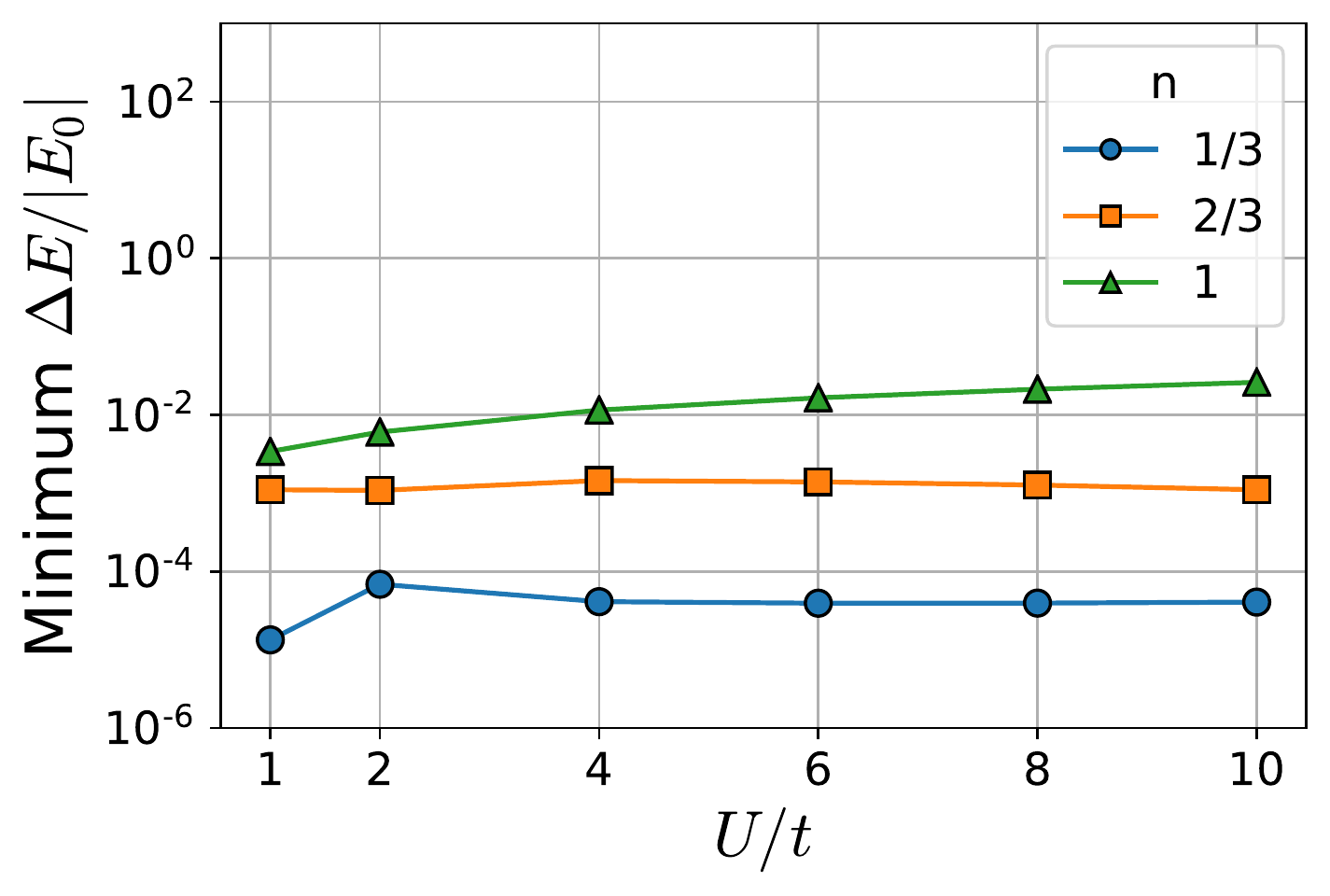}
    \label{fig:hubbard_min_delE_diff}
    }\\
    \caption{Minimum $\Delta E/|E_0|$ out of all checkerboard pattern drivers on a $4\times 3$ FHM, between the annealed state and the true $H_{LW}$ ground state at various doping $n$ and interaction strength $U$. All drivers had no terms on the Auxiliary qubits.
    Annealing simulation was performed with a linear schedule and total time $T=100$ in arbitrary units. 
    In (a), the checkerboard pattern for both spins match, where in (b) spin up/down have different drivers. For matching drivers, we show the addition of 2 and 3 $H^{XX}$ Heisenberg driver terms in dotted lines in (a) which greatly improves the fidelity. }
    \label{fig:hubbard_min_delE}
\end{figure}

In Fig.~\ref{fig:hubbard_min_delE} we show the corresponding energy error to those in Fig.~\ref{fig:hubbard_max_fid}. We find that despite the wide range of fidelity reported in the main text, we obtain low energy errors. For example, for matching spin drivers at $U/t=10$ at $n=2/3$ with no $H^{XX}$ terms, the annealing fidelity is $\approx 0.65$ with an energy error of $\approx 0.02$. 

\section{Spectral gap}\label{sec:app_sp_gap}
\begin{figure}
    \centering
        \subfloat[]{
      \includegraphics[width=0.5\columnwidth]{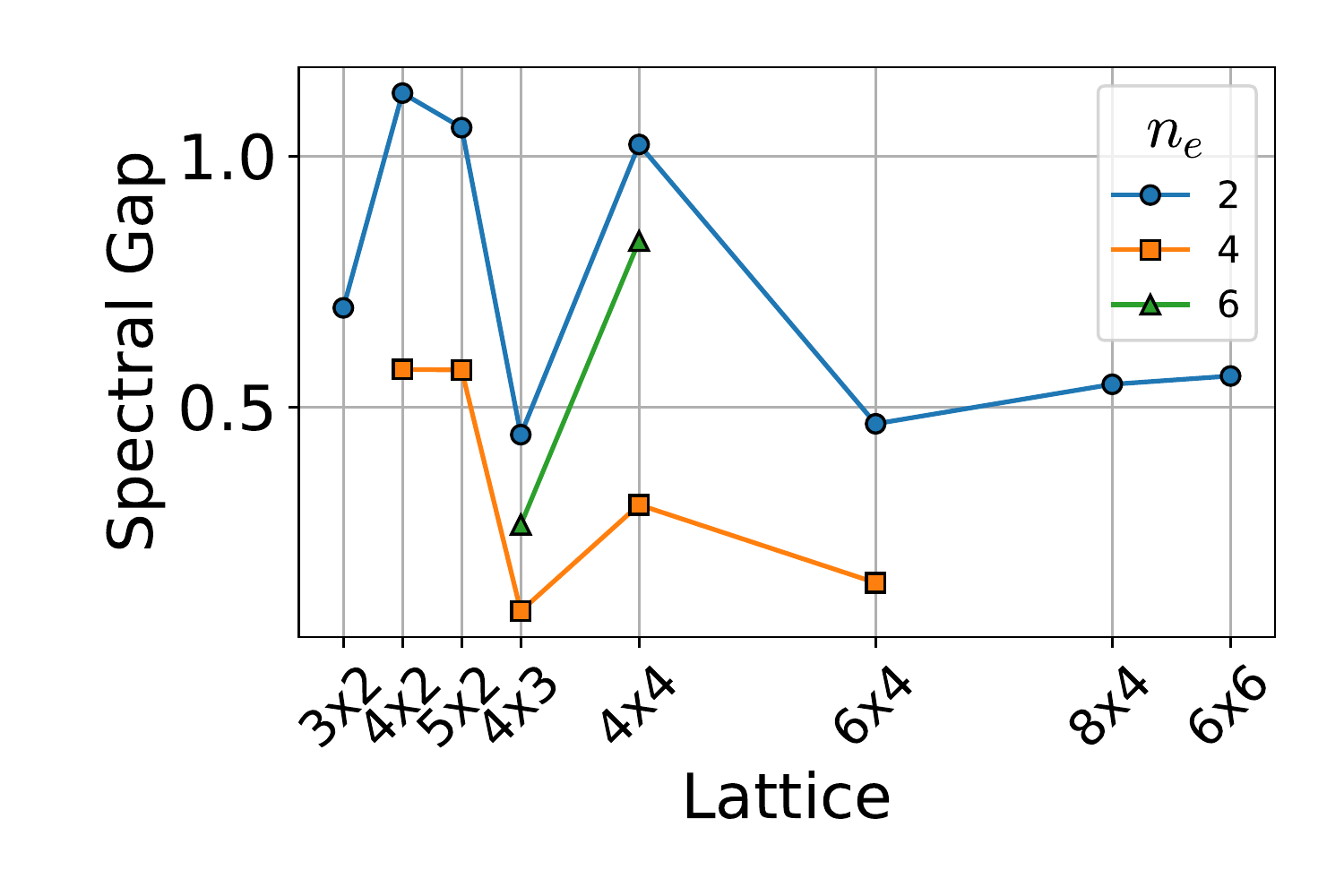}
    \label{fig:gap_spectral_spinless}
    }
    \subfloat[]{
      \includegraphics[width=0.5\columnwidth]{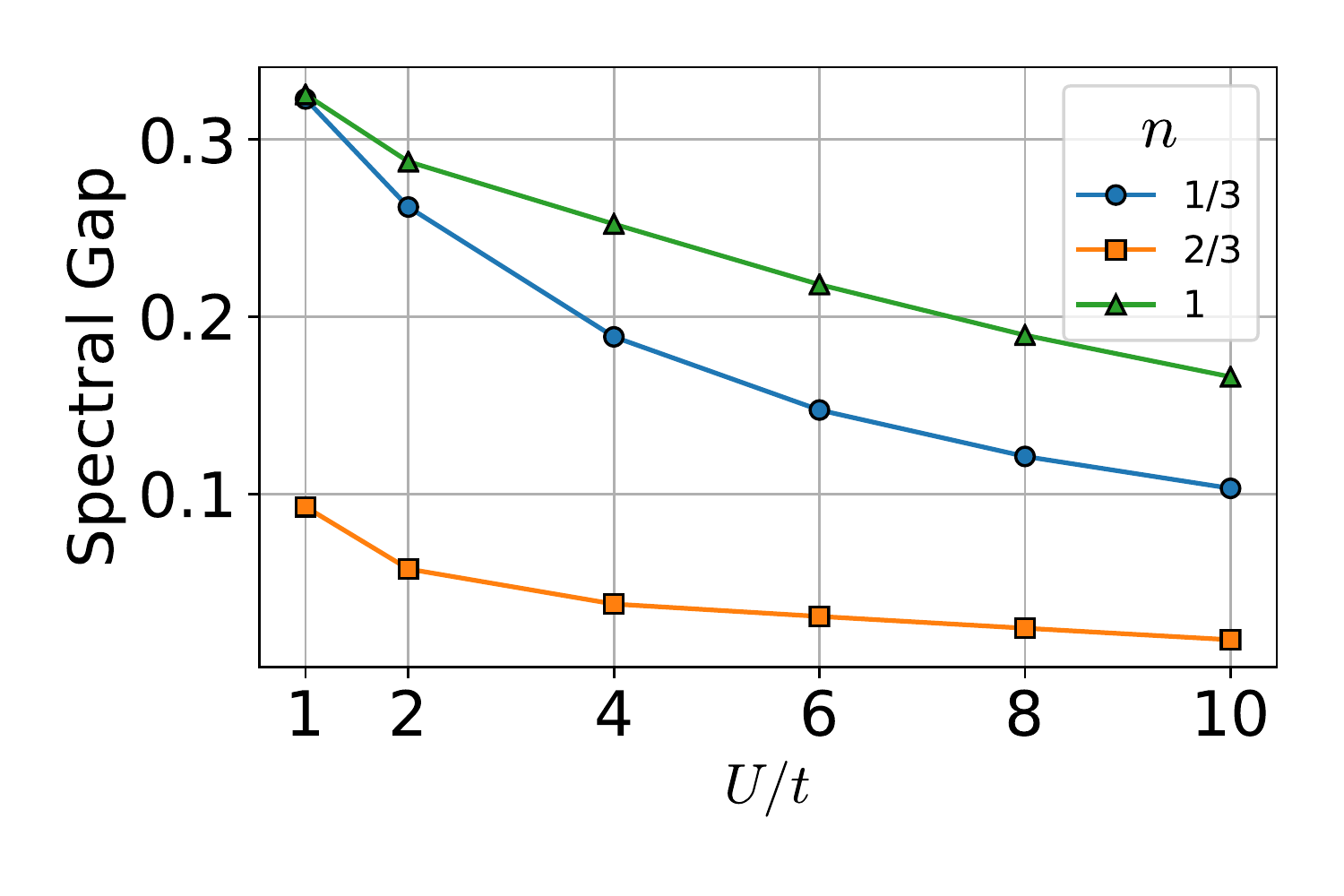}
    \label{fig:gap_spectral_hubb}
    }
    \caption{
    a) spectral gap $E_1 - E_0$ of the spinless $t-V$ model for various lattice sizes and number of electrons with $V/t=1$.
    b) spectral gap $E_1 - E_0$ for the $4 \times 3$ Hubbard model.
    }
    \label{fig:gap_spectral_all}
\end{figure}

We show in Fig.~\ref{fig:gap_spectral_all} the spectral gaps of the considered systems. The spinless system appears to have large fluctuations in value between system sizes, while the Hubbard (spinful) systems show a slow decay as interaction strength is increased.

% \begin{figure}
%     \centering
%     \subfloat[]{
%       \includegraphics[width=0.5\columnwidth]{gap_spectral_spinless.pdf}
%     \label{fig:gap_spectral_spinless}
%     }
%         \subfloat[]{
%       \includegraphics[width=0.5\columnwidth]{gap_abs_spinless.pdf}
%     \label{fig:gap_abs_spinless}
%     }
%     \caption{a) spectral gap $E_1 - E_0$ of the spinless $t-V$ model for various lattice sizes and number of electrons with $V/t=1$.
%     b) absolute minimum gap $\min_s E_1(s)-E_0(s)$ for various lattice sizes and number of electrons alongside the spectral gap from (a). The drivers correspond to those in Fig.~\ref{fig:spinless_deltaE}.  }
%     \label{fig:gap_spinless_all}
% \end{figure}

% \begin{figure}
%     \centering
%     \subfloat[]{
%       \includegraphics[width=0.5\columnwidth]{gap_spectral_hubb.pdf}
%     \label{fig:gap_spectral_hubb}
%     }
%         \subfloat[]{
%       \includegraphics[width=0.5\columnwidth]{gap_hubbard.pdf}
%     \label{fig:gap_hubbard}
%     }\\
%     \caption{a) spectral gap $E_1 - E_0$ for the $4 \times 3$ Hubbard model.
%     b) relative spectral gap as a function of total anneal $s = \frac{t}{T}$ for the $4\times 3$ Hubbard model at $n=2/3$ doping. The drivers correspond to those in Fig.~\ref{fig:hubbard_max_fid_same} i.e. matching spin drivers. The minimum gap occurs at the end of the anneal, corresponding to the spectral gap. }
%     \label{fig:gap_hubbard_all}
% \end{figure}

\section{Hubbard Noise Study} \label{sec:app_noise}
  
\begin{figure}
    \centering
       \includegraphics[width=0.5\columnwidth]{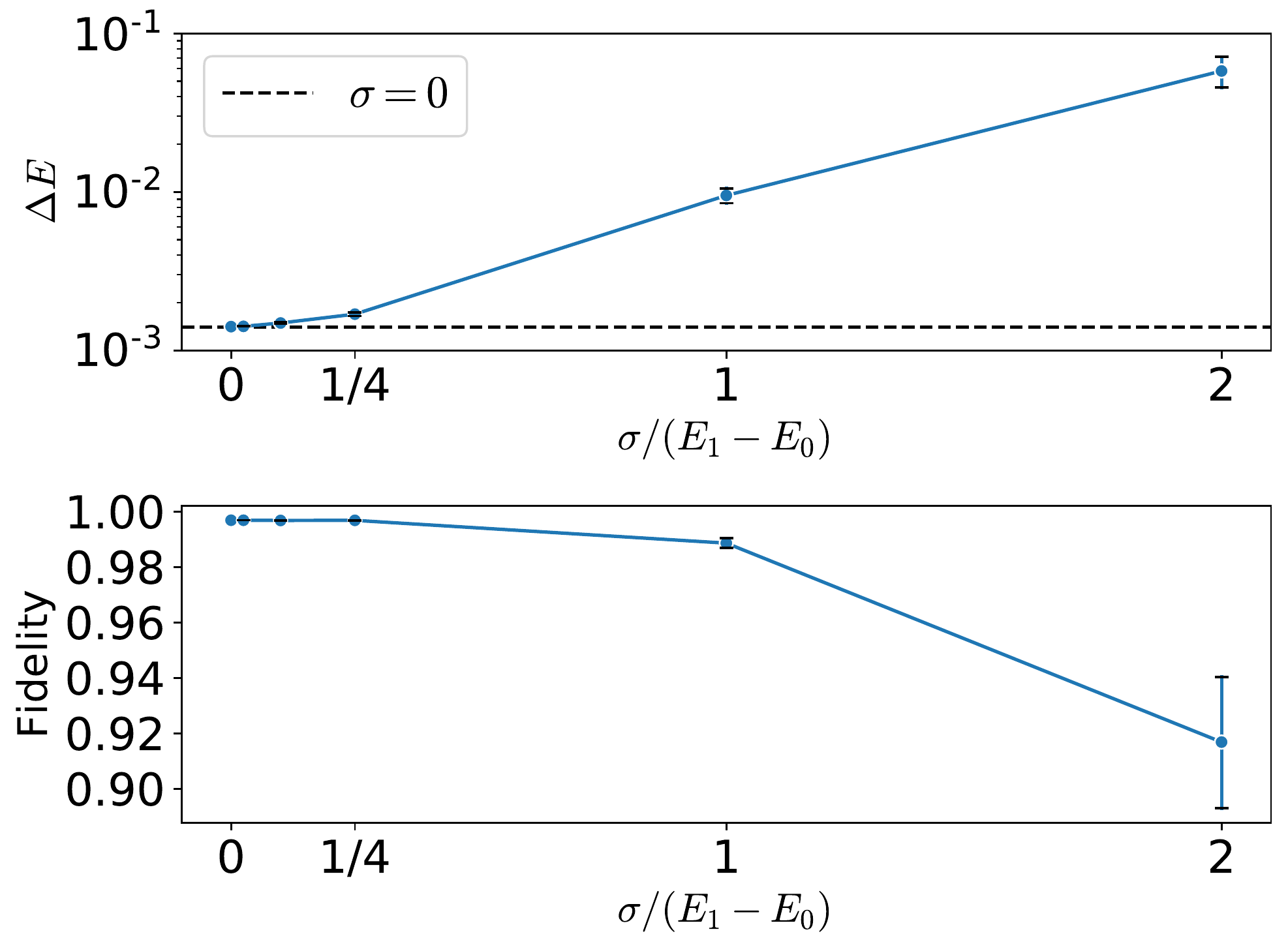}
    \caption{
    \textit{top}: Average energy difference $\Delta E - \langle H_{LW} \rangle - E_0$ and \textit{bottom}: fidelity when adding random Gaussian noise to each $H_i$ representing control errors for a $4\times 3$ Hubbard model at $n=2/3$ and $U/t=4$. Results are averaged over 50 random realizations. Noise is centered around $t=1$ or $U=4$ with a variety of width $\sigma $, given in units of the spectral gap $E_1-E_0 \approx 0.04$.
    The drivers correspond to those in Fig.~\ref{fig:hubbard_max_fid_same}.  }
    \label{fig:app_noise_results}
\end{figure}

\begin{figure}
    \centering
       \includegraphics[width=0.75\columnwidth]{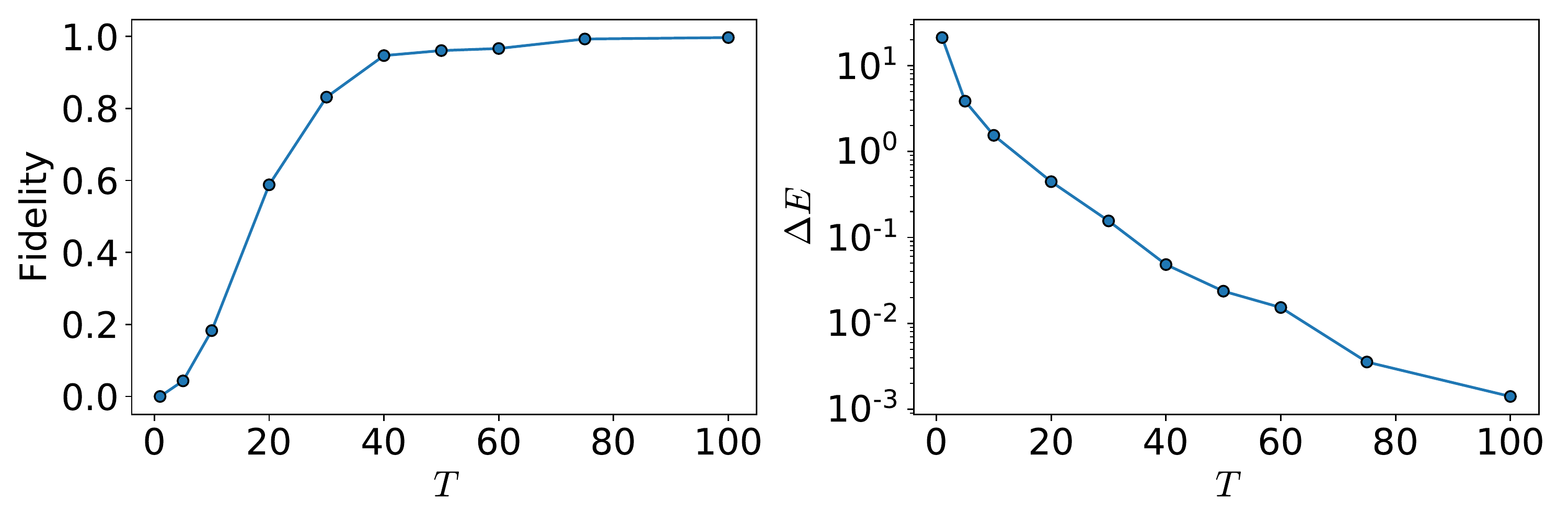}
    \caption{
    Fidelity and energy difference $\Delta E - \langle H_{LW} \rangle - E_0$ for various total annealing times $T$ for a $4\times 3$ Hubbard model at $n=2/3$ and $U/t=4$.
    The driver correspond to those in Fig.~\ref{fig:hubbard_max_fid_same}.  }
    \label{fig:app_timing_results}
\end{figure}

To understand the robustness of the protocol in the logical subspace to implementation noise and limitations, we focus on a single point in the phase diagram: the Hubbard model at $U/t=4$ at $n=2/3$ doping on a $4\times 3$ lattice. First, we consider control errors within the logical subspace. The LW Hamiltonian can be broken down into $4+1$ non-commuting terms $H_1,\dots,H_5$ shown as a cartoon in Fig.~\ref{fig:model_FH}e. For each of these terms, we consider a Gaussian perturbation centered around the expected value of $t=1$ or $U=4$ with standard deviation $\sigma$. Similar methodology has been used to study D-Wave quantum annealers \cite{Brugger2022}. 

We can simulate the quantum annealing process for this symmetry broken Hamiltonian and compare with the noise-free solution, shown in Fig.~\ref{fig:app_noise_results}. For low noise compared to the spectral gap, which is $E_1-E_0\approx 0.04$ for this interaction strength and doping, we find little to no performance degradation. When increasing the noise to be on the order of the spectral gap, the average energy error is increased about 6.7X but the fidelity only decreases from $\approx 0.996$ to on average $\approx 0.988(1)$. Thus we can speculate that if device control errors are in kept below the spectral gap we can maintain similar performance as the noise-free case. We leave a more comprehensive understanding of the noise as well as its role in scaling to future work.

\end{document}